\documentclass{aa}
\usepackage{graphicx}
\usepackage{natbib}
\begin{document}

\title{Numerical computation of isotropic Compton scattering}
\titlerunning{Compton scattering}

\author{R. Belmont\inst{1,2}\thanks{\email{belmont@cesr.fr}} }
\institute{CESR (Centre d'\'Etude Spatiale des Rayonnement) ; Universit\'e de Toulouse [UPS] , 9 Av. du Colonel Roche, 31028 Toulouse, France \and 
CNRS ; UMR5187, 31028 Toulouse, France}
%\institute{Centre d'Etude Spatiale des Rayonnements (OMP; UPS; CNRS), 9 avenue du Colonel Roche, BP44346, 31028, Toulouse Cedex 4, France}

\date{Received --- / Accepted ---}

\abstract{Compton scattering is involved in many astrophysical situations. It is well known and has been studied in detail for the past fifty years. Exact formulae for the different cross sections are often complex, and essentially asymptotic expressions have been used in the past. Numerical capabilities have now developed to a point where they enable the direct use of exact formulae in sophisticated codes that deal with all kinds of interactions in plasmas. Although the numerical computation of the Compton cross section is simple in principle, its practical evaluation is often prone to accuracy issues. These can be severe in some astrophysical situations but are often not addressed properly.}
{In this paper we investigate numerical issues related to the computation of the Compton scattering contribution to the time evolution of interacting photon and particle populations.}
{An exact form of the isotropic Compton cross section free of numerical cancellations is derived. Its accuracy is investigated and compared to other formulae. Then, several methods to solve the kinetic equations using this cross section are studied.}
{The regimes where existing cross sections can be evaluated numerically are given. We find that the cross section derived here allows for accurate and fast numerical evaluation for any photon and electron energy. The most efficient way to solve the kinetic equations is a method combining a direct integration of the cross section over the photon and particle distributions and a Fokker-Planck approximation. Expressions describing this combination are given.}{}

\maketitle
\section*{Introduction}

Compton scattering has become a major process in astrophysics, and for more than half a century, it has been applied to a large variety of astrophysical situations, particularly in high energy, low density media. It was first introduced in the context of electron cosmic rays interacting with thermal stellar photons \citep[etc.]{F47,FP48,D51}.
It was then used to model gases of high energy electrons in the strong radiation field of X-ray or relativistic sources \citep{L71}. It is also responsible for the up-scattering of low energy photons from the microwave background radiation \citep{Z72}. More recently, much effort has been spent on high energy media in relativistic sources such as $\gamma$-ray bursts, AGN, and X-ray binaries. Particular focus is now on the synchrotron self-Compton radiation that is thought to play a major role for example in AGN \citep{Tavecchio01,Sauge04}. %Compton scattering is now included in most fields of modern astrophysics.

The many varied applications result in different energy regimes for the photons and the particles. The energy of particles spans from the very cold electrons of pulsars winds to the hot gas in the corona of X-ray binaries (E$\approx$ 100 keV, i.e. $E/m_ec^2\approx 1$), and to high energy cosmic rays (up to $E > 100$ TeV, i.e. $E/m_ec^2 >10^{8} $). The energy of photons also spans a wide range: in the case of AGN for instance, it goes from low energy synchrotron photons ($\lambda = 1$m, i.e. $h\nu/m_ec^2 \approx10^{-12}$) to $\gamma$-ray emission (with E $>10$ TeV, i.e. $h\nu/m_ec^2 >  10^7$ for some sources). General formulae must be able to deal simultaneously with very small and very large energies.

The exact cross section is quite complex and does not give any physical insight into the fundamental process. Moreover, often, only a fraction of the particle and photon spectra contribute significantly to the Compton scattering in one given astrophysical situation. For these reasons most of the theoretical work has been done on deriving limiting and/or averaged forms that can be easily included in the modeling of astrophysical objects. 
For example, much work is now based on the diffusion approach for particles, valid when large angle scattering can be neglected, that is, when the energy of the particles does not vary significantly in one single interaction \citep[work initiated by][]{K57}. Analytical expressions averaged over Maxwellian distributions and thus only valid in purely thermal plasmas have also been used. Many other approximations have been proposed in general papers, reviews and books \citep[see][for example]{BG70,Z75,RL79,K87,N94}. Nevertheless the relevant energy cannot always be confined to a narrow range where simple approximations can be made. Also more general expressions that can be used regardless of a specific problem are required for generic codes designed to address various situations. In these cases an exact description without limitation on the photon and electrons energies is required. A few exact formulae have been proposed for the isotropic Compton cross section \citep[NP94 hereafter]{Jones68, Brinkmann84, N94}. Such formulae are analytically equivalent. When evaluated numerically however, they behave differently with respect to accuracy issues. For example, the original formula by \citet{Jones68} fails to describe the up-scattering of low energy photons which represents a typical astrophysical situation. Although these problems are well known, the Jones' cross section is still widely used in the literature.

Such general expressions are not very helpful for physical understanding. Nevertheless, they are complementary to asymptotic forms for they provides quantitative results that can be compared to observations. It has become particularly true recently, since more and more codes have been developed to deal with photon-photon, photon-particle and particle-particle interactions in high energy plasmas \citep[e.g.][]{Coppi92, Poutanen96, NM98, Belmont08, Poutanen08}. Even with an accurate, exact cross section, numerical computation of the Compton contribution to the evolution of the particle and photon distributions is problematic. When the distributions and cross section are discretized in energy bins, the grid resolution puts severe constraints on the code accuracy. Typically, high energy particles are scattered down by numerous and inefficient scattering events off soft photons. If the resolution is insufficient, these individual events are not captured by the numerical scheme and the global evolution of the particle distribution is not described accurately. \citet{NM98} proposed that a combination of this method with a Fokker-Planck approximation could enable good accuracy. This idea was applied in a recent numerical code by \citet{Belmont08} \citep[see also][]{Poutanen08}. Here we investigate the method and quantify its accuracy.

In the first section, we derive a new form of the distribution of scattered photons. The accuracy of this formula is discussed and compared to other expressions. In the second section, numerical errors of several kinetic methods are estimated and compared.  

\section{Evaluating the isotropic distribution of Compton-scattered photons}

%To keep models simple, such general scattered distribution is often described by its first moments, typically the first 3: the total number of scattered photons, their mean energy and the dispersion\footnote{Note that in the diffusion limit this is sufficient.} \citep{Coppi90,N94}. Although not as complicated as the general cross section, exact formulae for the first moments are already quite complex and hard to evaluate numerically, so that, limiting forms are generally used in turn. In addition, describing the distribution by its first 3 moments is not accurate when it is not symmetrical (for low energy particles for instance). At this point, a general expression for the scattered distribution of photons must be considered. 

\subsection{Exact spectrum}
\label{spectrum1}

As long as stimulated scattering is not considered, the evolution of a photon population interacting by Compton scattering with an electron distribution is described by the following equation \citep[see][for example]{RL79}:
\begin{eqnarray}
&&\frac{\partial f_\omega(\vec{k}) }{\partial t} =  \int  d\vec{p}_0  f_e(\vec{p}_0)  \int d\vec{k}_0  \times\\
&&\left[     f_\omega(\vec{k}_0)  c \frac{ d\sigma}{d\vec{k}} (\vec{p}_0,\vec{k}_0 \rightarrow \vec{k}) \right. \nonumber 
 -
\left. f_\omega(\vec{k})  c\frac{d\sigma}{d\vec{k}_0}(\vec{p}_0, \vec{k} \rightarrow \vec{k}_0) 
 \right]   ~.
\end{eqnarray}
Here $f_{\omega}=d{\cal N}_{\omega}/d^3\vec{r}/d^3\vec{k}$ and  $f_{e}=d{\cal N}_{e}/d^3\vec{r}/d^3\vec{p}$ are the photon and lepton distributions in their respective 6-dimension momentum space. The first term in the brackets describes the contribution of all photons of momentum $\vec{k}_0$ being scattered to momentum $\vec{k}$ after one interaction. The second one describes photons of momentum $\vec{k}$ being scattered to any other momentum. Both are proportional to the Compton differential cross section $d\sigma/d\vec{k}(\vec{p}_0,\vec{k}_0 \rightarrow \vec{k})$ giving the probability of one photon of momentum $\vec{k}_0$ being scattered to a momentum $\vec{k}$ by an electron of momentum $\vec{p}_0$. This cross section can be interpreted as the distribution of scattered photons resulting from the interaction of mono-energetic photons with one single electron and it will be often be referred to as such hereafter.

When the photon and particle distributions are isotropic, this equation can be integrated over all directions. By noting $d^3\vec{p}/(m_ec)^3=p^2 dp d \vec{\Omega}_p$ and $d^3\vec{k}/(m_ec)^3=\omega^2 d\omega d \vec{\Omega}_\omega$ (where $\omega=h\nu/m_ec^2$), it yields:
\begin{eqnarray}
\frac{\partial N_\omega(\omega)}{\partial t}  &=& \int \hspace{-.3cm} \int  c \frac{d\bar{\sigma}}{d\omega}(p_0,\omega_0\rightarrow\omega) dN_\omega(\omega_0)d N_e(p_0)    \nonumber \\ 
  &&  \quad \quad \quad \quad- N_\omega(\omega)   \int  c\bar{\sigma}_0(p_0,\omega_0)   d N_e(p_0)  \label{eq2}
\end{eqnarray}
where $N_{\omega}(\omega) = 4\pi  \omega^2 f_\omega(\vec{k})$ and $N_p(p) = 4\pi  p^2 f_p(\vec{p})$ are the angle-integrated distributions.
Here
\begin{equation}
\bar{\sigma}_0(p_0,\omega_0) = \int \frac{d \bar{\sigma}}{d\omega}(p_0,\omega_0\rightarrow\omega) d\omega 
\end{equation}
is the total scattering cross section, and
\begin{equation}
\frac{d\bar{\sigma}}{d\omega}(p_0,\omega_0\rightarrow\omega) =\int \frac{d\vec{\Omega}_{e,0}}{4 \pi} \int \frac{d\vec{\Omega}_{\omega,0}}{4\pi}\int \frac{d\sigma}{d\omega d\vec{\Omega}_\omega} d\vec{\Omega}_\omega \label{average}
\end{equation}
is the angle-averaged cross section for isotropic Compton scattering. 

\begin{figure}[h!]
\begin{center}
\includegraphics[scale=.4]{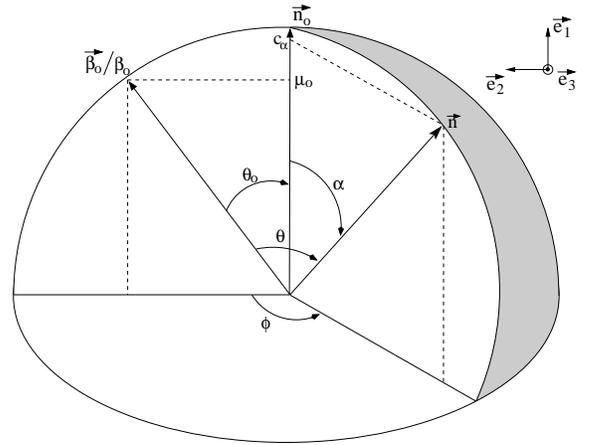}
\caption{Angles and notations. The incoming photon direction $\vec{n}_0$ is defined by the incident angle $\theta_0$ and an azimuthal angle $\psi_0$ that orientates the global picture (not shown here). The scattered photon direction $\vec{n}$ is defined by the scattering angle and the azimuthal angle $\phi$. In coordinates $(\vec{e}_1,\vec{e}_2,\vec{e}_3)$, $\vec{\beta}_0/\beta_0=(\mu_0,\sin{\theta_0},0)$ and $\vec{n}=(c_\alpha,\sin{\alpha}\cos{\phi},\sin{\phi})$, so that $\mu = \mu_0 c_\alpha+ \sin{\alpha}\sin{\mu_0}\cos{\phi}$.}  \label{angles}
\end{center}
\end{figure}

Eq. \ref{average} can actually be simplified further. Because of symmetry, the differential cross section $d\sigma/d\omega/d\vec{\Omega}_\omega$ does not depend independently on the direction of the incoming photon and lepton, but only on the angle between their directions: $\mu_0 = \cos{\theta_0}$, $\theta_0=(\vec{\beta}_0,\vec{n}_0)$ where $\beta_0=p_0/\gamma_0$ is the lepton velocity in units of the speed of light and $\vec{n}_0$ is the direction of the incoming photon (see Fig. \ref{angles}).
Then, Eq. \ref{average} reduces to:
\begin{equation}
\frac{d\bar{\sigma}}{d\omega}(\vec{p}_0,\omega_0\rightarrow\omega) =\int_{-1}^{1} \frac{d\mu_0}{2} \int \frac{d\sigma}{d\omega d\vec{\Omega}_\omega} d\vec{\Omega}_\omega ~. \label{average1}
\end{equation}
Also, although the result of a single scattering event is described by 6 variables (energy and direction of the scattered photon and particle), these variables are not independent. The conservation of the total energy-momentum 4-vector sets 4 constrains, so that only 2 independent variables are necessary to describe the scattering outcome. Although other choices can be made, these are often chosen to be the angles describing the scattered photon direction $\vec{n}$: the cosine of the scattering angle $c_\alpha=\cos{\alpha}$, $\alpha=(\vec{n}_0,\vec{n})$ and an azimuthal angle $\phi$ (see Fig. \ref{angles}). Then, the differential cross section $d\sigma/d\omega/d\vec{\Omega}_\omega$ is redundant and includes a delta-function that ensures that the total energy-momentum is conserved:
\begin{equation}
\frac{d\bar{\sigma}}{d\omega}(\vec{p}_0,\omega_0\rightarrow \omega) =\frac{1}{2} \int \frac{d\sigma}{dc_\alpha d\phi} \delta(\omega - \bar{\omega}) d\mu_0 d\phi dc_\alpha~. \label{average1}
\end{equation}
Here the latter condition is given by:
\begin{equation}
\bar{\omega} = \omega_0 \frac{1-\beta_0\mu_0}{1-\beta_0\mu +\frac{\omega_0}{\gamma_0}(1-c_\alpha)}
\end{equation}
where $ \mu=\mu_0c_\alpha + \sqrt{1-\mu_0^2}\sqrt{1-c_\alpha^2} \cos{\phi}$ is the cosine of the angle between the direction of the scattered photon and the incoming electron. 

The angle differential cross section is a well known result of quantum mechanics. In the rest frame of the electron, the general Klein-Nishina cross section does not depend on the azimuthal angle and reads \citep{H54,RL79}:
\begin{equation}
\frac{d^2 \sigma'}{d\phi' dc_\alpha'} = \frac{1}{2\pi} \frac{d\sigma'}{dc_\alpha'}= -\frac{3 \sigma_T}{16 \pi}  \rho'^2  \left( \rho' + \frac{1}{\rho'} -1 +c_\alpha'^2 \right)  \label{sigmaERF}
\end{equation}
where $\sigma_T$ is the Thomson cross section, $\rho'(c_\alpha) = \omega'/\omega'_0 = 1/(1+\omega'_0(1-c_{\alpha}'))$ is the ratio of the scattered to incoming photon energy. Here, all variables denoted with ' refer to quantities in the rest frame of the particle.

Eq. \ref{average1} and \ref{sigmaERF} can be shown to be equivalent to Eq. 12 of \citet{Jones68} and Eq. 6.2.1, 8.1.1 and 8.1.3 of NP94. Getting the angle-averaged cross section is now quite straightforward. However, including the Klein-Nishina cross section in Eq. \ref{average1} requires changes of frame and the triple integral quickly leads to cumbersome expressions. Different integration variables can be used and the triple integral can be performed in several orders. Here we follow the original method by \citet{Jones68}. All quantities are expressed in the electron rest frame which gives:
\begin{equation}
\frac{d\bar{\sigma}}{d\omega}=  \int  \frac{d\sigma'}{dc_\alpha'} \frac{\delta\left[ \omega-\bar{\omega}\right]}{\gamma_0^4(1+\beta_0\mu_0')^3}\frac{d\mu_0'}{2} \frac{d\phi'}{2\pi} dc_\alpha'~. \label{sec_diff}
\end{equation}
Integration is then performed over $\mu_0'$, $c_\alpha'$ and then $\phi'$. The latter integration is made with the variable $1+\beta_0\mu'(\phi')$. The resulting formula by \citet{Jones68} contains several misprints\footnote{Contrary to what is claimed in \citet{Coppi90}, all intermediate expressions are correct and only the final equations 23-27 contains misprints. }. Using slightly different notations, the correct formula for the scattered distribution reads \citep[see also][]{PW05}:
\begin{equation}
\frac{d\bar{\sigma}}{d\omega}(p_0,\omega_0;\omega) = \frac{3\sigma_T}{8\gamma_0p_0\omega_0^2}\left[ F(x_-,x_+) \right]_{a}^{b}
\end{equation}
with 
\begin{eqnarray}
F  &=& \frac{L_+}{\omega \omega_0}  \\
& + & \frac{1}{2L_+}\left( x_--\frac{1}{x_-} -4  -(\gamma/\omega+\gamma_0/\omega_0) \right)   \nonumber \\ 
 & + &\frac{1}{2L_-}\left( x_--\frac{1}{x_-}  - 4 + (\gamma_0\omega_0+\gamma\omega) \right)      \nonumber \\
& + & \frac{c_+}{\lambda_+}\left( \frac{1}{L_+} -   {\rm acosh} \left(\sqrt{\lambda_+x_+/2}\right)/\sqrt{\lambda_+}  \right)   \nonumber  \\ 
& +& \frac{c_-}{\lambda_-} \left( \frac{1}{L_-} - \left\{ 
\begin{array}{l}
{\rm asin~~} \left(\sqrt{|\lambda_-| x_-/2}\right)/\sqrt{|\lambda_-|} ~ \rm{if} ~ \lambda_-<0 \\
{\rm asinh} \left(\sqrt{|\lambda_-| x_-/2}\right)/\sqrt{|\lambda_-|} ~~\rm{if}~ \lambda_->0 
\end{array}
\right. \right)  \nonumber \label{Eq1}
\end{eqnarray}
where $\gamma_0$ and $\gamma=\gamma_0+\omega_0-\omega$ are the Lorentz factors of the incoming and scattered particles respectively and $p=\sqrt{\gamma^2-1}$ is the momentum of the scattered particle.
Here the following notations have been used:
\begin{eqnarray}
\lambda_+ &=& p^2_0 + 2\gamma_0\omega_0 + \omega_0^2\\
 \lambda_- &=& p_0^2 -2\gamma_0\omega~ + \omega^2  \\
L_\pm &=& \sqrt{\lambda_\pm \mp 2/x_\pm}  \\
c_\pm &=& 1+\omega\omega_0+2\lambda_\pm ~.
\end{eqnarray}
The cross section is proportional to the difference $\left[F\right]_a^b$ of function $F$ evaluated at the upper and lower boundaries of the last integration ($b$ and $a$ respectively). At these boundaries, the variables $x_\pm$ have the following expressions:
\begin{eqnarray}
x_+^a = {\rm min}\left[(\gamma_0+p_0 )/\omega_0 ;  (\gamma+p)/\omega~ \right] &,& x_-^a = 1/(\omega\omega_0x_+^a) \\
x_-^b = {\rm min}\left[(\gamma_0+p_0)/\omega ~; (\gamma+p)/\omega_0 \right] &,& x_+^b = 1/(\omega\omega_0x_-^b) ~.
\end{eqnarray}
The critical photon frequencies for which the arguments of the ${\rm min}$ and ${\rm max}$ functions are equal define distinct domains of the scattered distribution. For the lower integration boundary $a$, both arguments are equal at a unique frequency: $\omega=\omega_0$. For the higher boundary, both arguments are equal at two critical frequencies: $\omega=\omega_0$ and $\omega=\omega_c$, where
\begin{equation}
\omega_c =  \omega_0 \frac{\gamma_0+p_0}{\gamma_0-p_0 + 2\omega_0} ~.
\end{equation}
The scattered distribution is obviously continuous at these frequencies. However, higher derivatives are not, as it can be seen in Fig. \ref{spectraA} for example.

The scattered distribution is bounded in energy and the maximal and minimal energies guarantee that all radicals remain real \citep[see][for more details]{Jones68}:
\begin{eqnarray}
\omega_{\rm min} &=&  \omega_0 \frac{\gamma_0-p_0}{\gamma_0+p_0 + 2\omega_0} \label{bdr1} \\
\omega_{\rm max} &=& \left\{ \begin{array}{ll} 
\omega_0+\gamma_0-1  &~~~ {\rm if }~~~ \omega_0 > \omega_0^*(p_0)  \\
\omega_c  & ~~~{\rm if }~~~\omega_0 < \omega_0^*(p_0)  \end{array} 
 \right.~, \label{bdr2}
\end{eqnarray}
where 
\begin{equation}
\omega_0^*=\frac{1}{2}(1+p_0-\gamma_0) ~. \label{omegaB}
\end{equation}
When, $\omega_0>\omega_0^*$, the photon can gain up to the entire electron energy. However, this absolute limit is not necessarily reached. When  $\omega_0<\omega_0^*$ (as for Compton up-scattering for instance), only a fraction of the electron energy is at most transfered to the photon.

\subsection{Cancelation issues}
As was first discussed by \citet{Jones68}, evaluation of Eq. \ref{Eq1} suffers from accuracy issues due to machine round-off errors. Since most astrophysical problems involve an energy range spanning many orders of magnitude, Eq. \ref{Eq1} includes combinations of very small and very large terms. In this form, most large terms must cancel out, which obviously represents a numerical challenge. An example of accuracy issue is shown in Fig. \ref{spectraA}.

Cancelation errors appear when the relative difference of two terms is smaller than the machine precision. Here we focus on computations in double precision, i.e. when reals are coded in 8 octets (64 bits) and their mantissa is coded in 52 bits. The relative error due to machine precision is then of the order of $\epsilon = 2^{-52} \approx 10^{-15}$. 

%First, the scattered distribution cannot be computed when $(\omega_{\rm max}-\omega_{\rm min})/\omega_{\rm min} < \epsilon $, that is approximatively when $p_0+2\omega_0 < \epsilon$, simply because the frequency $\omega$ can not be evaluated accurately enough. However, 
%the distribution is then so peaked at $\omega=\omega_0$ that there is little point computing the exact distribution. 

\begin{figure}[h!]
\begin{center}
\includegraphics[scale=.5]{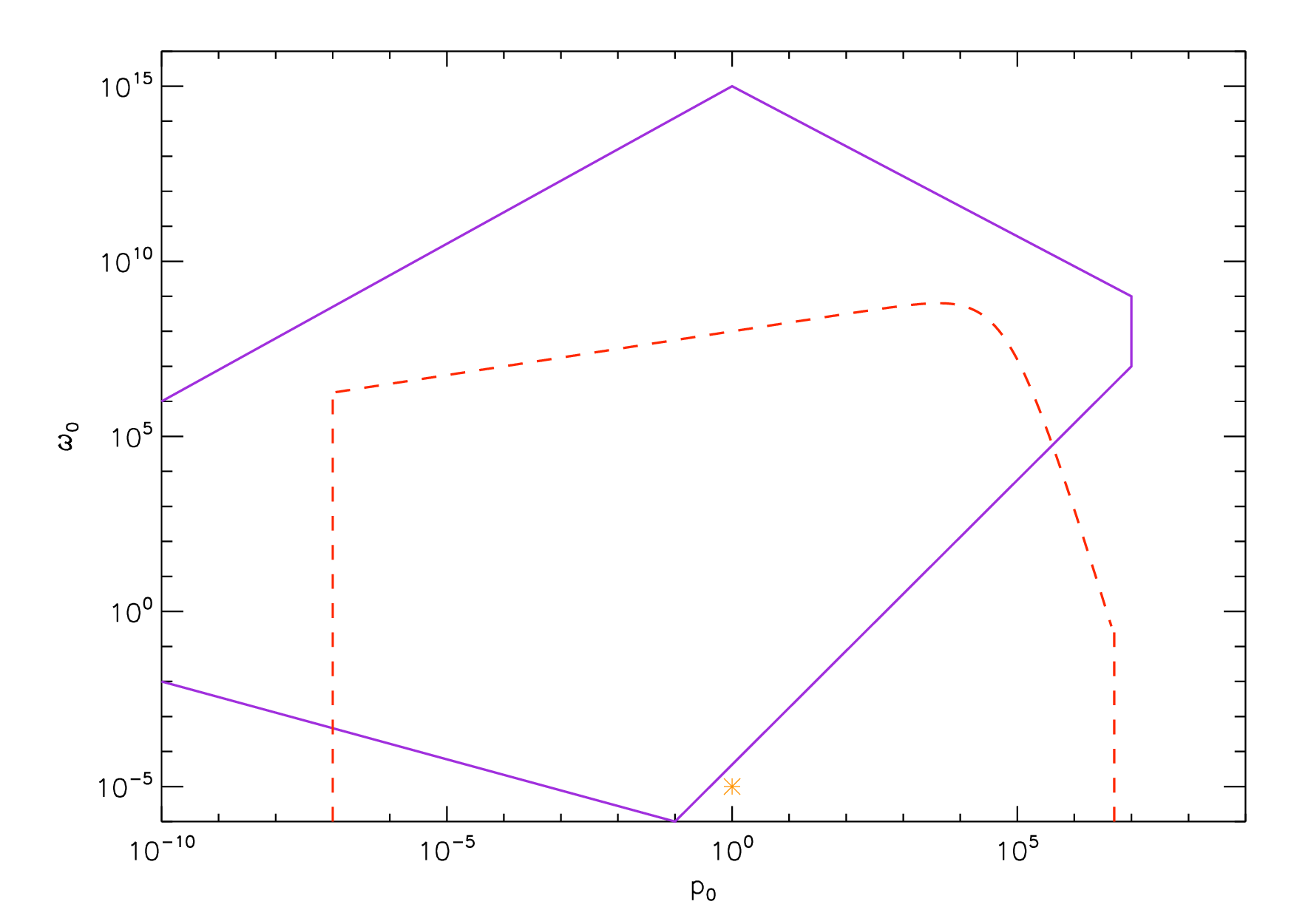}
\caption{Accuracy domains. Equation \ref{Eq1}  is accurate for photons energy and electron momentum $(\omega_0,p_0)$ in the region delimited by the solid line. The original formula derived by \citet{Jones68}, and the cross section by \citet{PW05}, have basically the same properties and are accurately evaluated in the same region. The published formula by NP94 is accurate in the domain determined by the dashed line. These boundaries are for double precision computing. The star shows the point for which the scattered distribution is plotted in Fig. \ref{spectraA}.}  \label{domain}
\end{center}
\end{figure}

We have checked the accuracy of the scattered distribution for a large range of photon and lepton energies and identify the domain of the $(\omega_0,p_0)$ plane where Eq. \ref{Eq1} gives accurate results. Results are shown in Fig. \ref{domain}. 
We find that Compton scattering of mild photons ($\omega_0\sim1$) by mid-relativistic particles ($p_0\sim 1$) is well described by this formula but problems arise for lower or higher energies. In particular, the scattered distribution cannot be evaluated in the most typical astrophysical case, namely the up-scattering of soft photons by high energy particles. 

In the latter case, \citet{Jones68} derived asymptotic expansions in $\omega_0/\gamma_0<<1$. However, a large number of orders must be kept when $\omega_0/\gamma_0\la1$ and the numerical evaluation often becomes time consuming. Moreover, other situations are also affected by numerical issues, for which such an expansion is not relevant. Here, we rather concentrate on analytical manipulation of the original formula to get an exact expression free of cancelation issues.

\subsection{Re-writing the exact cross section}
\label{spectrum2}
Numerical accuracy issues result from two kinds of cancelations: when the relative difference of function $F$ evaluated at the two integration boundaries is very small and when some terms constituting $F$ should cancel out. Here we deal with both successively.

\subsubsection{Cancelation issues when $|F^b-F^a| << F^a, F^b$}
This happens typically when $x_\pm^a\approx x_\pm^b$, as for example, at the distribution upper and lower photon energies $\omega_{\rm min}$ and $\omega_{\rm max}$ where the distribution vanishes, but where $F^a$ and $F^b$ can remain large.

Differences of hyperbolic functions between the upper and lower boundaries can be computed analytically by using the following trigonometric relations: 
\begin{eqnarray}
{\rm acosh} (\sqrt{a}) - {\rm acosh} (\sqrt{b}) &=& {\rm asinh}\left( \sqrt{a(b-1)}-\sqrt{b(a-1)} \right) \nonumber \\
{\rm asinh} (\sqrt{a}) - {\rm asinh} (\sqrt{b}) &=& {\rm asinh}\left( \sqrt{a(1+b)}-\sqrt{b(1+a)} \right) \nonumber \\
{\rm asin~} (\sqrt{a}) - {\rm asin~} (\sqrt{b}) &=& {\rm asin ~}\left( \sqrt{a(1-b)}-\sqrt{b(1-a)} \right) \nonumber
\end{eqnarray}
The difference of the two hyperbolic functions between the two integration boundaries $a$ and $b$ simply reads: ${\rm asinh}(\Delta \sqrt{\lambda_\pm})$  if $\lambda_\pm >0$, or ${\rm asin}(\Delta\sqrt{-\lambda_\pm} )$ if $\lambda_\pm<0$, where
\begin{eqnarray}
\Delta &=& L_+(x_+^b)-L_+(x_+^a) \\
      &=&  L_-(x_-^b)-L_-(x_-^a) \\
      &=&{\rm min}\left[p,p_0\right] +  {\rm min}\left[ 0 ;  \omega+\omega_0-p-p_0 \right]/2 ~.
\end{eqnarray}

Although the last term of Eq. \ref{Eq1} is not divergent it is numerically ill-behaved when $\lambda_- << 1$ and it is best to use the following auxiliary function:
\begin{eqnarray}
S(x) &=& \left\{ \begin{array}{c} 
\frac{1}{x} \left( \frac{{\rm asinh} (\sqrt{x})}{\sqrt{x}} -\frac{1}{\sqrt{1+x}} \right) {~\rm if~} x>0 \\
\frac{1}{x} \left( \frac{{\rm asin} (\sqrt{-x})}{\sqrt{-x}} -\frac{1}{\sqrt{1+x}} \right) {~\rm if~} x<0 
\end{array} \right. 
\end{eqnarray}
which is evaluated as a series expansion for small argument:
\begin{equation}
S(x) = \sum_{n=0}^{\infty} \frac{(2n+1)!(-x)^n}{4^n (n!)^2(2n+3)} ~~{\rm for} ~~|x|<1~. \label{DL_S}
\end{equation}

Differences of all other terms in Eq. \ref{Eq1} between the two integration boundaries can also be computed analytically to factorize $\Delta$ and get the following quite simple expression for the scattered distribution:
\begin{equation}
\frac{d\bar{\sigma}}{d\omega} = \frac{3 \Delta}{8\omega_0^2\gamma_0p_0} G \label{Eq2_0}
\end{equation}
with
\begin{eqnarray}
G  &=& 2 + (1+\omega\omega_0)\left[ z_+ + z_--\frac{2}{\omega\omega_0}\right] \nonumber \\
    &+& 2 \left[A \right]_-^+ - \Delta^2(1+\omega\omega_0)\left[zA\right]_-^+ \nonumber \\
    &+& \Delta^2\left[ cz^{3/2}S\left(\lambda \Delta^2 z\right)\right]_-^+  \label{Eq2}
\end{eqnarray}
where $ A_\pm = (1/z_\pm+\lambda_\pm \Delta^2)^{-1/2} $ and $z_\pm=x_\pm^ax_\pm^b$.
The scattered distribution is proportional to $\Delta$. This factor holds as the difference between the two integration boundaries $a$ and $b$ and vanishes at the minimal and maximal photon energies $\omega_{\rm min}$ and $\omega_{\rm max}$. For high particle energy, it is best computed as:
\begin{eqnarray}
\Delta &=& {\rm min} \left\{ {\rm min}(p;p_0)~;~ 
\frac{\gamma_0+\gamma+p_0+p}{2(p+p_0)} \right. \nonumber \\
 & & \quad \quad \quad \times \left. \left( {\rm min}(\omega;\omega_0) - \frac{{\rm max}(\omega;\omega_0)}{(\gamma+p)(\gamma_0+p_0)} \right) \right\} ~.
\end{eqnarray}

Here we note that an accurate evaluation of the scattered electron momentum $p$ is crucial. When the latter is low (typically when $p<10^{-3}$), the simple derivation $p=\sqrt{(\gamma_0+\omega_0-\omega)^2-1}$ is not accurate enough and an alternative derivation must used:
\begin{equation}
p = p_0 - \frac{(\omega-\omega_0)(p+p_0)(\gamma+\gamma_0)}{2p_0(p+p_0)-(\omega-\omega_0)(\gamma+\gamma_0)}~.
\end{equation}

Equations \ref{Eq2_0}-\ref{Eq2} are accurate over a larger domain than the original formula and can be used in many astrophysical situations.

\subsubsection{Cancelation when $|z_+-z_-| << z_+,z_-$}
However, this new expression reveals other differences hidden in the original one and that result in cancelation issues, namely when $|z_+-z_-| << z_+,z_-$. In such cases, the differences must be computed analytically. Although alternate expressions for the differences can often be accurate in more cases, it is sufficient to use them when the relative differences are smaller than $\epsilon=10^{-10}$.

%First, the momentum of the scattered electron is not well estimated by $p=\sqrt{\gamma^2-1}$ when its value become extremely low (typically $p<10^{-14}$). Then it is better estimated by:
%\begin{equation}
%p = p_0 + \frac{(\omega_0-\omega)(\gamma+\gamma_0)(p_0+\sqrt{\gamma^2-1})}{2p_0(p_0+\sqrt{\gamma^2-1})+(\omega_0-\omega)(\gamma+\gamma_0)}
%\end{equation}

By using the definition of $z_+$ and $z_-$, the differences in the first three terms of Eq. \ref{Eq2} can be computed as:
\begin{eqnarray}
 z_+ &+& z_--\frac{2}{\omega\omega_0} = \nonumber \\ 
  & & \left( \frac{1}{\omega}-\frac{1}{\omega_0}\right)^2 \mbox{min}\left[1,\frac{(\gamma+p+\gamma_0+p_0)^2\omega\omega_0}{(\gamma_0+p_0)(\gamma+p)(p+p_0)^2}\right] 
\end{eqnarray}
\begin{eqnarray}
\left[A \right]_-^+ &=& \frac{-A_-^2A_+^2}{A_++A_-} (\Delta^2 \left[\lambda\right]_-^+ - \omega^2\omega_0^2 \left[z\right]_-^+ ) \\
\left[zA\right]_-^+ &=& \frac{z_++z_-}{2}\left[A \right]_-^+  + \frac{A_++A_-}{2}\left[z\right]_-^+ 
\end{eqnarray}
respectively, where
\begin{eqnarray}
\left[ \lambda\right]_-^+ &=& (\omega+\omega_0)(\gamma+\gamma_0) \\
\left[z\right]_-^+ &=& -\left|\frac{1}{\omega}-\frac{1}{\omega_0}  \right| \\
&& \times \mbox{min}\left[ \frac{(\gamma+p+\gamma_0+p_0)^2}{(p+p_0)(\gamma+p)(\gamma_0+p_0)},\frac{1}{\omega}+\frac{1}{\omega_0} \right] ~.
\end{eqnarray}
In the same manner, the last term can be computed as:
\begin{equation}
\left[ cz^{3/2}S\right]_-^+ = \frac{S_++S_-}{2}\left[cz^{3/2}\right]_-^+ + \frac{c_+z_+^{3/2}+c_-z_-^{3/2}}{2} \left[S\right]_-^+
\end{equation}
where
\begin{eqnarray}
\left[cz^{3/2}\right]_-^+ &=& \frac{z_+^{3/2}+z_-^{3/2}}{2}\left[ \lambda\right]_-^+  \\
 && +\frac{c_++c_-}{2} \left[ z\right]_-^+\frac{1+z_+/z_-+z_-/z_+}{z_+^{1/2}/z_-+z_-^{1/2}/z_+} ~.
\end{eqnarray}
The numerical cancelation of $\left[S \right]_-^+$ fails either when $\lambda_+ z_+ \approx \lambda_- z_- $ or when $\Delta^2 \lambda_\pm z_\pm << 1$ and these two cases must be dealt with separately. In the former case, first order Taylor developments of $S$ around the mid value $\Delta^2 (\lambda_+ z_++\lambda_- z_-)/2$ are sufficient whereas in the latter case (typically when $\Delta^2 \lambda_\pm z_\pm << 10^{-2} $), series expansions of $S$ in zero can be used (Eq. \ref{DL_S}).  In both cases, the difference is proportional to $\left[\lambda z \right]_-^+$ which must be computed as:
\begin{equation}
\left[\lambda z \right]_-^+ = \frac{z_++z_-}{2}\left[\lambda\right]_-^+ + \frac{\lambda_++\lambda_-}{2}\left[z\right]_-^+ 
\end{equation}
for low photon energy ($\omega_0 << 1$). With the additions of this subsection, Eq. \ref{Eq2} is accurately evaluated  for any particle and photon energy.

\subsection{Comparisons to other formulae}
\label{validity}
\begin{figure}[h!]
 \begin{center}
 \includegraphics[scale=.53]{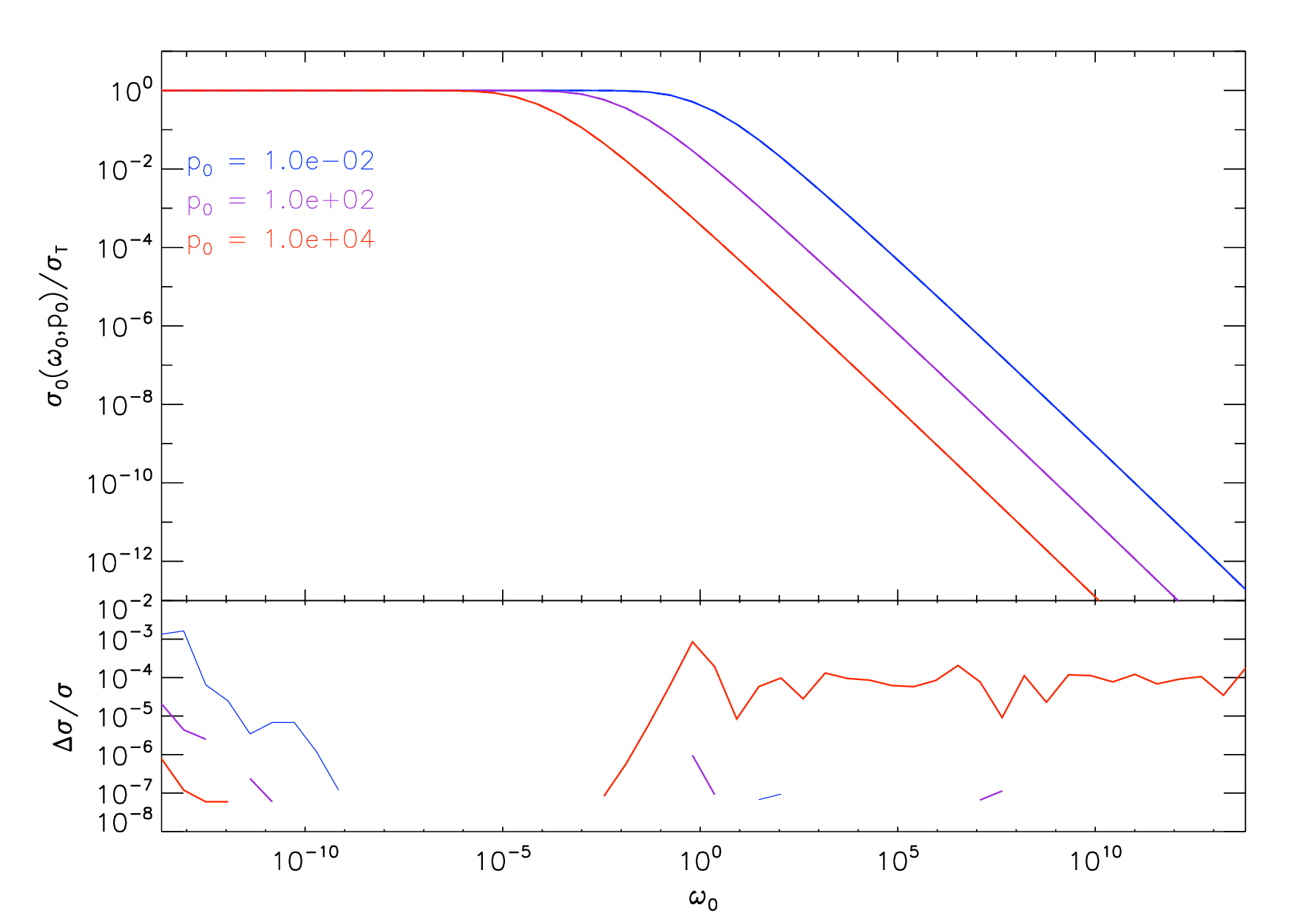} 
\includegraphics[scale=.53]{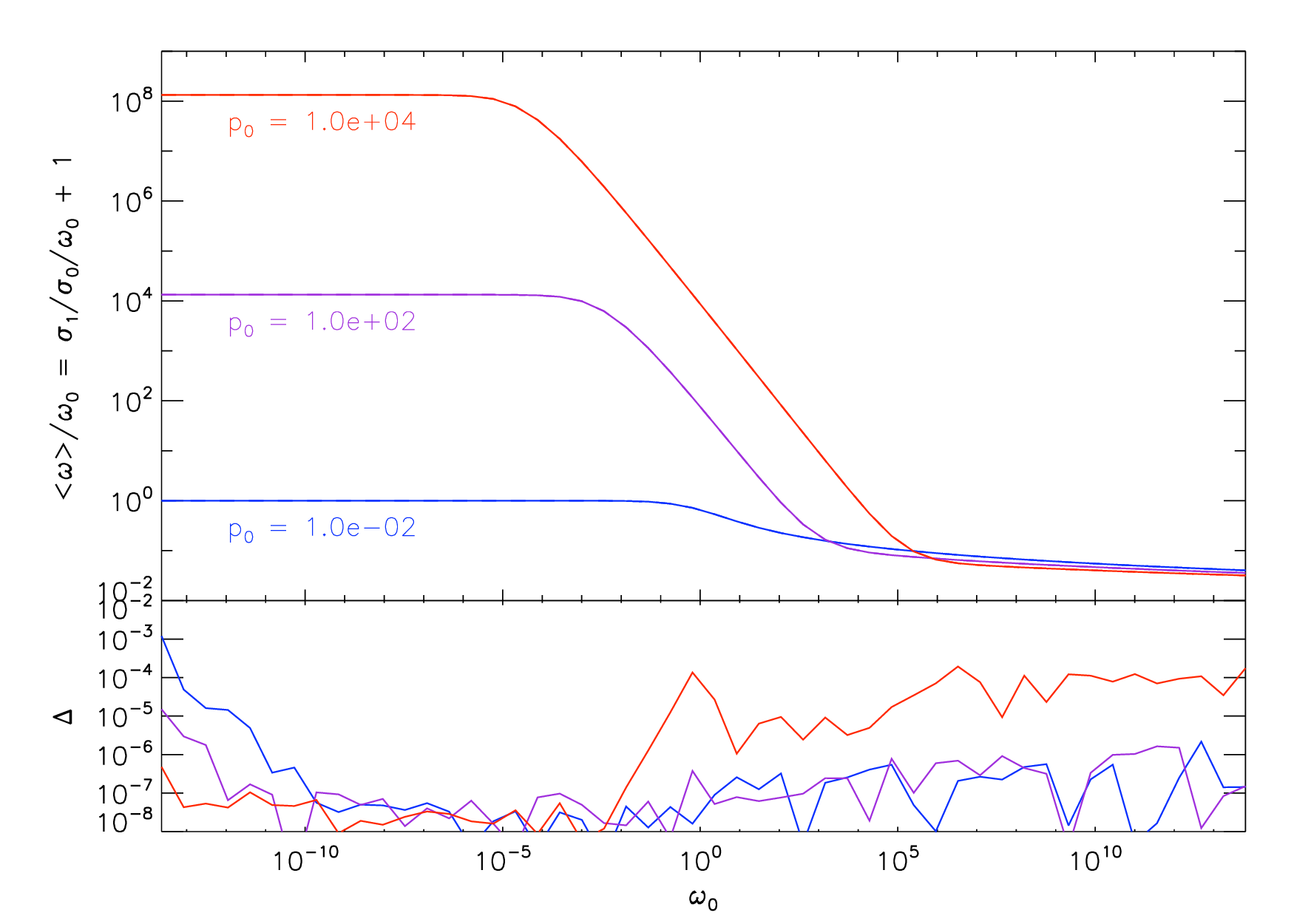} 
\includegraphics[scale=.53]{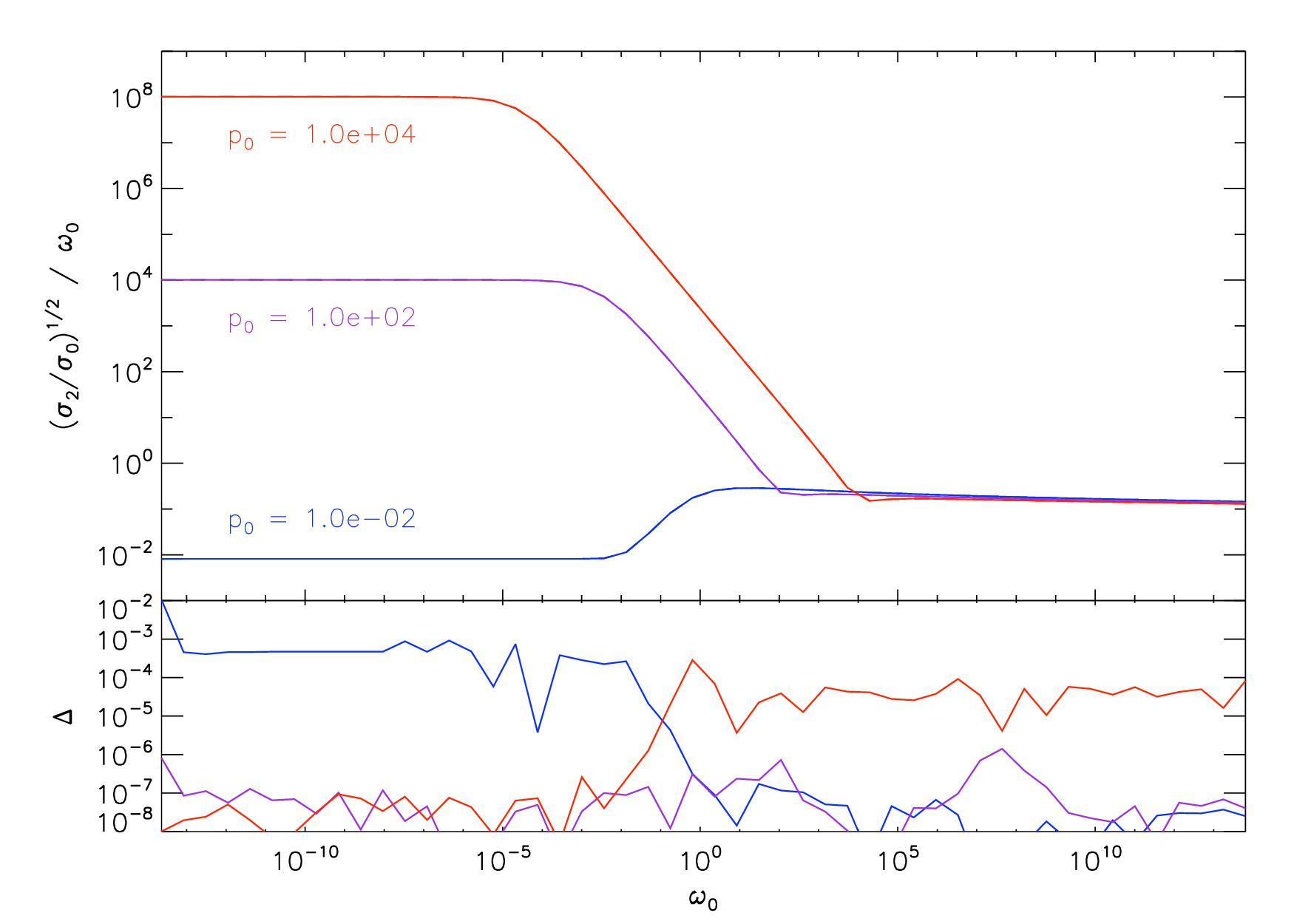}
\caption{First three moments of the scattered distribution as a function of the energy of the incoming photon: the total cross section, the mean energy of the scattered photon and the dispersion about this mean value. The solid and dashed lines show the numerical integration of the expression derived in this paper and analytical results given in NP94 respectively. The 3 curves are for $p_0=10^{-2},10^2$ and $10^4$. For each panel, relative errors are also shown.} \label{moment0}
 \end{center}
 \end{figure}

This expression was checked extensively against previous formulae. The exact differential cross section was first compared to the original Jones' formula (identical to \ref{Eq1} when the misprints are corrected) and to the formula by NP94\footnote{The equations in NP94 contain some misprints too. In particular: Eq. 8.1.10 should read $\mu_+ = 1- (x-x_1)^2/D_m/x/x_1$ and Eq. 8.1.22 should read $H_1 = \left[ H_0-\Delta h A_1(h_-) + H/A(h_-)A(h_+)\right]/h_+ $} (Eq. 8.1.7-8.1.22). The latter was derived by integrating Eq. \ref{sec_diff} in a different order and behaves numerically better than the Jones' cross section for Compton up-scattering of soft photons (see Fig. \ref{domain}).  Results show perfect agreement of the three formulae when the numerical evaluation of the three cross sections is accurate. The first moments of the distribution:
\begin{eqnarray}
\sigma_0(p_0,\omega_0) &=& \int \frac{d\bar{\sigma}}{d\omega} d\omega \\
\sigma_1(p_0,\omega_0) &=& \int (\omega-\omega_0) \frac{d\bar{\sigma}}{d\omega} d\omega \\
\sigma_2(p_0,\omega_0) &=&\int (\omega-\omega_0)^2 \frac{d\bar{\sigma}}{d\omega} d\omega
\end{eqnarray}
 - namely the total cross section, the mean energy and the dispersion of the scattered photon energy - were also computed numerically by integrating over the scattered distribution and they were compared to analytical equations\footnote{A few misprints were corrected. In Eq. 3.3.5, $\Psi_{14}$ should read $(2\psi_{-11}-\psi_{-10}-\psi_{-12})/\xi$, the last expression for Eq. 3.3.9 should be $\pi^2/6+\ln^2{(2\xi)}/2 - g_*(1/2\xi)$, and in Eq. 3.3.10, $\psi_{00}$ should read: $3/(8\xi)\left[ g(\xi)-2/\xi+ (1/2+2/\xi+1/\xi^2)l_\xi-R_\xi/2-3/2\right]$.} (NP94, Eq. 3.3.1-3.1.10). Examples are shown in Fig. \ref{moment0}. 
Again, good agreement is obtained. Small differences are observed but they result from numerical errors in the integration of the scattered distribution. 

\begin{figure}[h!]
\begin{center}
\includegraphics[scale=.53]{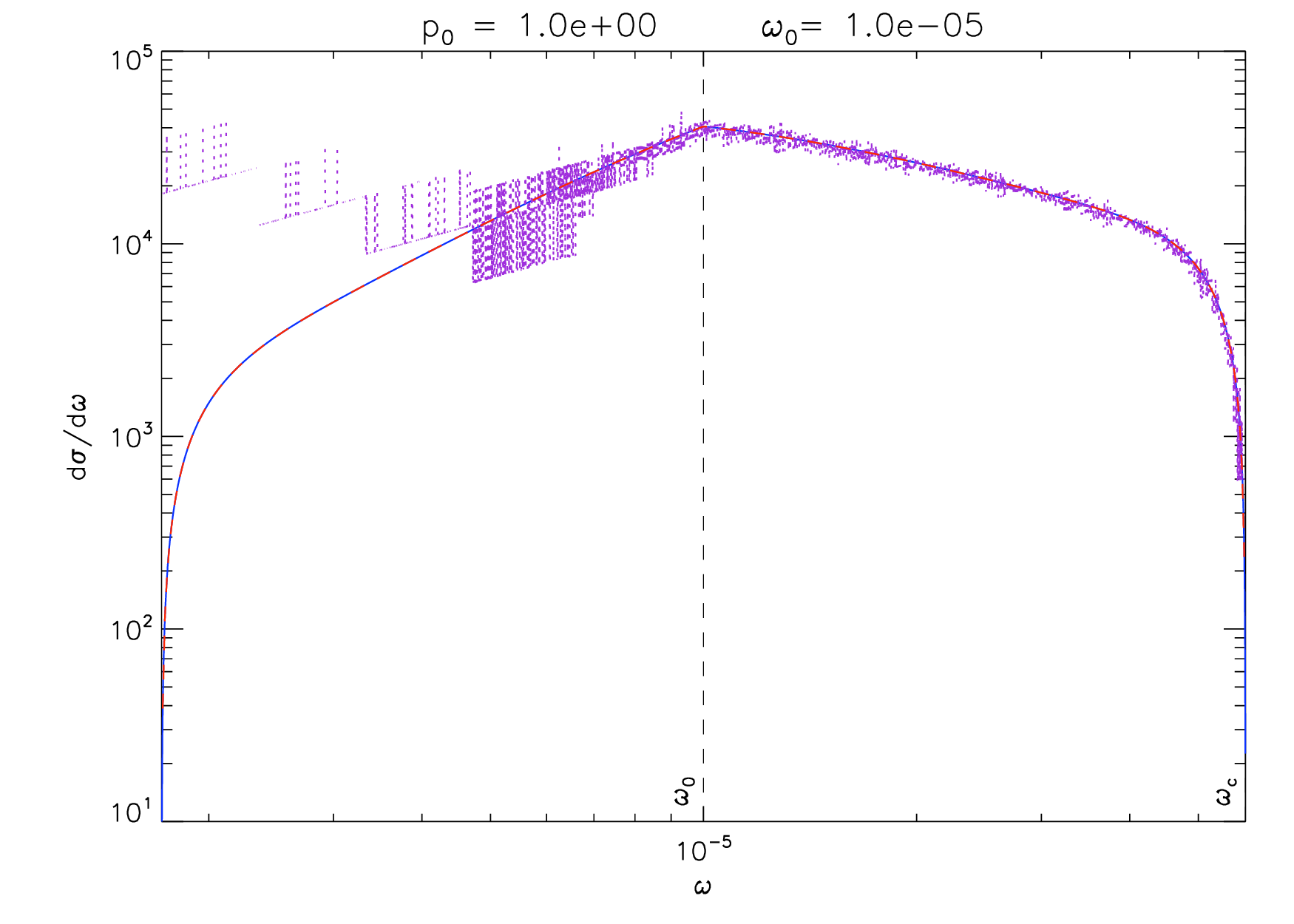} 
\caption{The scattered distribution for soft photons ($h\nu=10^{-5}m_ec^2$) and mild relativistic particles ($p=1$). Solid and dotted lines show the distributions obtained with Equation \ref{Eq2}, and with Jones' formula respectively.} \label{spectraA}
\end{center}
\end{figure}

Compared to previous ones, the cross section derived here is found to be accurate over a wider energy range. An example of the scattered distribution is shown in Fig. \ref{spectraA}. As summed-up in Fig. \ref{domain}, the original cross section by \citet{Jones68} is inaccurate in typical astrophysical situations involving low energy photons up-scattered by high energy particles. The cross section derived by NP94 is accurate in most astrophysical cases. When computed directly from the published equations, it fails, however, for very high energy particles or very high energy photons. Although these cases may not be physically relevant since the Klein-Nishina cross section drops at these high energies, numerical issues can generate infinite values, propagate and affect numerical solutions. Nevertheless, it must be noted that a few changes in the way some quantities are computed enable good accuracy over a much larger domain (J. Poutanen, private communication). The cross section derived here is accurate for all photon and particle energies and overcomes these numerical issues.

%\begin{figure}[h!]
%\begin{center}
%\includegraphics[scale=.5]{Pw1}
%\includegraphics[scale=.5]{Pw2}
%\includegraphics[scale=.5]{Pw3}
%\caption{ Same as Fig. \ref{spectraA}. On the upper panel ($p_0=10^{-8}, \omega_0=10^8$), the formula by NP94 fails and the scattered distribution cannot be seen. }  \label{spectraB}
%\end{center}
%\end{figure}

%\clearpage
\section{Solving the kinetic equations}
Once the Compton cross section is computed accurately, the Compton contribution to the evolution of interacting particles and photons is described by the set of equations:
\begin{eqnarray}
\partial_t N_{e}(p) &=&  \int\hspace{-.3cm} \int c \frac{d\bar{\sigma}}{dp}(p_0,\omega_0 ; \omega(p))   dN_{e}(p_0) dN_\omega(\omega_0)   \nonumber  \\ 
                                                    & & \quad \quad  \quad \quad -  N_{e}(p) \int c \bar{\sigma}_0(\omega_0,p) d N_\omega(\omega_0) \label{compt_e_int} ~, \\
\partial_t N_\omega(\omega) &=&  \int \hspace{-.3cm} \int c \frac{d\bar{\sigma}}{d\omega}(p_0,\omega_0 ; \omega)  dN_e(p_0) dN_\omega(\omega_0)  \nonumber \\ 
                                                    & & \quad \quad  \quad \quad -  N_\omega(\omega) \int c \bar{\sigma}_0(\omega,p_0) dN_e(p_0) ~,  \label{compt_nu_int} 
\end{eqnarray}
where $\omega(p) = \omega_0 +\gamma_0-\gamma$ is the energy of the scattered photon and $d\bar{\sigma}(p_0,\omega_0;\omega(p))=d\bar{\sigma}(p_0,\omega_0;\omega)$ is the differential cross section of the interaction $(p_0,\omega_0)\rightarrow(p,\omega)$. In this section we show that computing these integrals numerically can also lead to accuracy issues. We present an alternative approach that combines this integration with a Fokker-Planck treatment and overcomes most numerical issues. The basic idea of such a combination was proposed by \citet{NM98} for particles only but was not investigated in detail. A more precise analysis of this treatment was presented in \citet{Belmont08} \citep[see also][]{Poutanen08}. Here we investigate the regimes where the various methods are valid and we focus on their numerical accuracy.

\subsection{Integral approach}
The simplest way to compute the time evolution of the lepton and photon populations is to discretize their distributions in energy bins. Integration over energy is then performed bin by bin by summing the contributions from all other bins. This process is time consuming since a double integral must be performed at each time step. Moreover it can be inaccurate in many situations:  when the width of the scattered distribution is smaller than the bin size, the energy change of the scattered particles (or equivalently photons) is too small to be resolved by the numerical scheme. The particles (or photons) do not scatter far enough in the energy space to reach other bins and there is no numerical evolution of the distribution. It is well known that most numerical issues arise from what happens below the grid scale. In this peculiar case however, particles (or photons) can undergo many scattering events per unit time when the density of scattering centres is large. Errors associated with individual scattering events add and they can lead to accuracy issues. This is typically what happens in simulations of X-ray binaries: the soft disc photons are up-scattered by the high energy particles constituting the corona. This interaction is known to efficiently cool the corona down. Whereas the photons experience large energy changes, the relative variation of the particle energy is very small and their cooling may not be captured by a discrete integral with finite resolution. The energy is then not conserved since the photon field gains energy while the particle population does not lose any\footnote{We note that the energy is not conserved only with non-linear grids. For linear grids and identical resolution in the photon and electron energy grids ($\omega_0$ and $\gamma_0$ respectively), the energy is balanced to machine precision. In such case indeed, the bin size is uniform. As the energy width of the scattered distribution is by definition the same for both species, if the scattering of particles to neighbour energy bins is not captured by the numerical integration, the scattering of photon is not either, so that the integral approach fails simultaneously for both species and there is no net energy exchange between the two species. Although the integral approach is clearly not accurate in this case, the energy balance is computed to machine precision.}. 

The accuracy of this integral method depends on the comparison of the energy bin size to the width of the scattered distribution. The latter depends on the momentum $p_0$ of the incoming particle and the energy $\omega_0$ of the incoming photon. Most astrophysical cases involve large ranges of energy (typically many orders of magnitudes) which implies the need to use logarithmic grids, the resolution of which is also energy-dependent. The accuracy of the numerical integration thus strongly depends on the region in the lepton momentum-photon energy space $(p_0,\omega_0)$ the simulation has to deal with. In this subsection, the regimes where the integral approach remains accurate are investigated. Hereafter the subscript 0 is dropped for simplicity.

\subsubsection{Photons}
Numerical integration for the evolution of the photon population is accurate in the limit where the width of the scattered distribution $\Delta\omega(p,\omega)=\omega_{\rm{max}}(p,\omega)-\omega_{\rm{min}}(p,\omega)$ is much larger than the local photon bin size $\delta \omega$. For linear grids, requiring that the scattered photon distribution spans at least $n$ bins would simply read\footnote{The $-1$ comes from the fact that when the width of the scattered distribution is exactly $n\delta\omega$, it is discretized in $n+1$ bins in all situations except when it is exactly centred at the centre of a bin.}: $\Delta\omega > (n-1)  \delta\omega$. For logarithmic grids, the same condition reads: $\omega_{\rm{max}}/\omega_{\rm{min}} > r_\omega^{n-1} $ where $r_\omega=\omega^{i+1}/\omega^{i} $ is the logarithmic increment of the photon grid. By denoting $R_\omega$ the number of bins per decade energy, i.e. the resolution of the photon grid, and $\epsilon_{\rm{I,\omega}}=10^{(n-1)/R_\omega}-1$, the condition for the integral approach to be valid reads: 
\begin{equation}
\Delta \omega(p,\omega)/\omega_{\rm{min}}(p,\omega)  >  \epsilon_{{\rm I},\omega} ~. \label{eq_bdr_I_photons}
\end{equation}
When the resolution is large enough: $\epsilon_{\rm{I},\omega} \approx 2 (n-1)/R_\omega$. However, as $n$ is typically of several, $(n-1)/R_\omega$ can be larger than unity for low resolution runs and the exact relation must be used. The condition \ref{eq_bdr_I_photons} is satisfied in a well defined region of the $(p,\omega)$ space. Using Eq. \ref{bdr1} and \ref{bdr2} to compute the scattered distribution width, one can write an equation in $\omega$ for the boundary of this region.  It is the solution of a 2nd order polynomial and the direct numerical integration is found to be accurate for all photon and particle energies {\it except} in the region where: 
\begin{equation}
 0 < p < p_{\rm I}^c \quad {\rm and}  \quad {\rm min}\left[\omega_{\rm I}^c(p),\omega_{\rm I}^-(p)\right] < \omega < \omega_{\rm I}^+(p) \\
\end{equation}
with
\begin{eqnarray} 
p_{\rm I}^c &=&  \frac{(\epsilon_{{\rm I},\omega}+4)  +(\epsilon_{{\rm I},\omega}-4)\sqrt{1+\epsilon_{{\rm I},\omega}}}{4\epsilon_{{\rm I},\omega}}  ~, \\
\omega^\pm_{\rm I} &=& \frac{1}{4} \left[ 2(1+p-2\gamma) + \epsilon_{{\rm I},\omega}(\gamma-p) \right.  \nonumber \\ 
  & \pm &  \left. \sqrt{\epsilon_{{\rm I},\omega}}\sqrt{4(\gamma-p-1)+\epsilon_{{\rm I},\omega}(1-2p\gamma+2p^2)} \right] ~, \\
\omega_{\rm I}^c&=&-\gamma + \frac{\epsilon/2}{(\gamma-p)\epsilon-2p} ~.
\end{eqnarray}
The inverse equation $p_{\rm I}(\omega)$ for this boundary would be more practical since it would be mono-valued and could be used directly  in the combined approach (see hereafter). It is the solution of a high order polynomial and is most easily found numerically. 

\begin{figure}[h!]
\begin{center}
\includegraphics[width=\columnwidth]{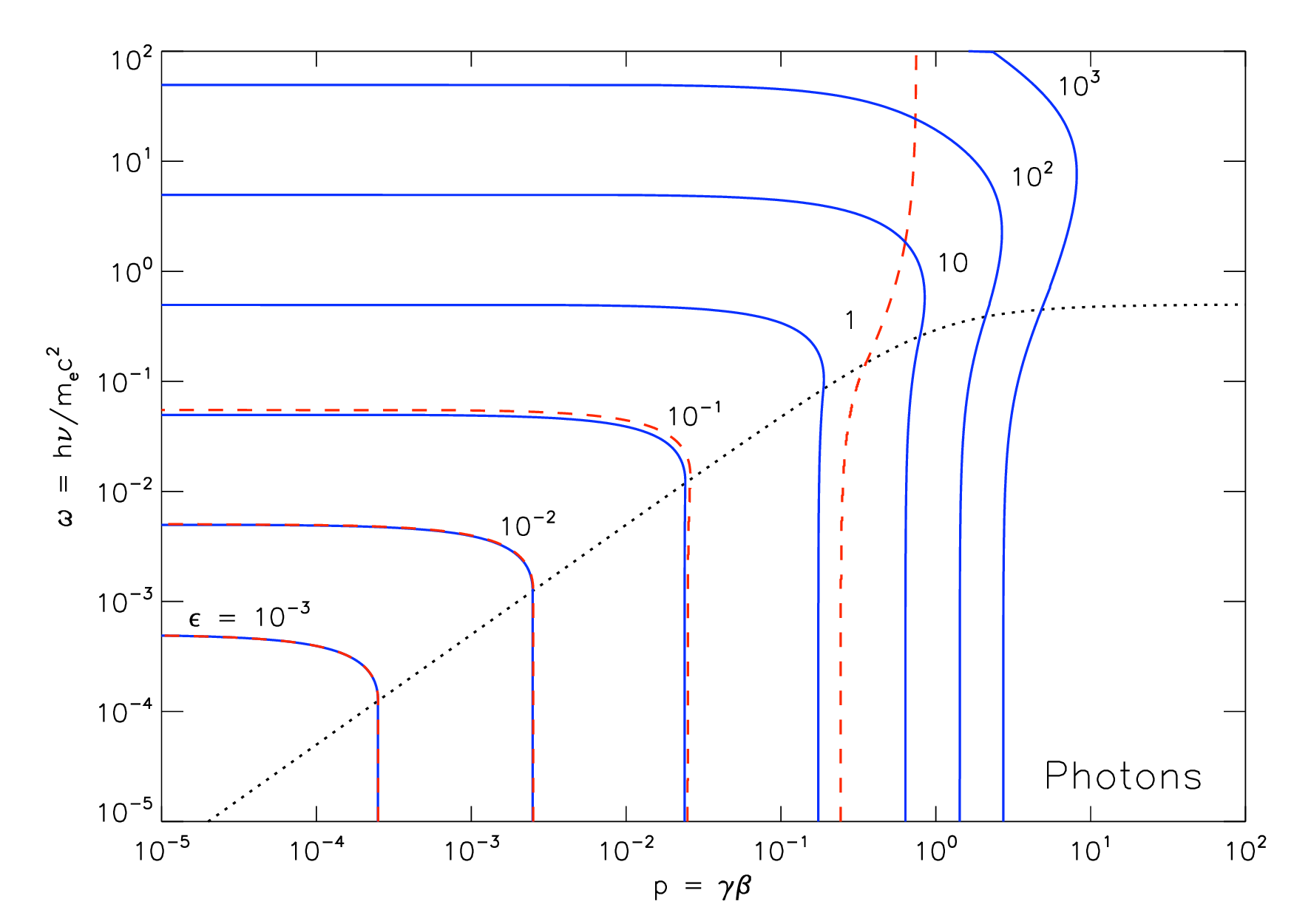}
\caption{Photon equation: boundaries $p_{{\rm I}}(\omega)$ for the integral approach (solid lines) and $p_{{\rm FP}}(\omega)$ for the Fokker-Planck approximation (dashed lines) for $\epsilon_{{\rm I},\omega}=10^{-3},10^{-2},10^{-1},1,10,10^2,10^3$ and $\epsilon_{\rm FP}=10^{-3},10^{-2},10^{-1},1$ respectively. The integral approach and the FP approximation are valid respectively right and left of the related boundaries. The dotted line shows $\omega_0^*(p)$ (see Eq. \ref{omegaB}).} \label{fig_photons}
\end{center}
\end{figure}

Boundaries for the integral approach are shown in Fig. \ref{fig_photons} for $R_\omega \approx 9200,930,97,13,3.8,2.0,1.3$ and $n=5$. For comparison typical runs have a resolution $R_\omega=1-30$. In the case $R_\omega=13$, only the scattering of low energy photons ($h\nu < 200$ keV) off sub-relativistic particles ($p < 0.2$) is inaccurately described by the integral approach. Although it is not general, the integral method for the photon equation can thus be used to address many of astrophysical situations, such as those involving high energy particles. 

\subsubsection{Particles}
Identically, the integral approach for particles is valid when the width of the scattered distribution is larger than the particle bin size. For logarithmic grids this condition is: $p_{\rm{max}}/p_{\rm{min}} > r_p^{n-1} $ where $p_{\rm{max/min}}=\sqrt{(\gamma+\omega-\omega_{\rm{min/max}})^2-1}$ are the largest and smallest momenta of the  scattered particle distribution and $r_p=p^{i+1}/p^i$ is the  logarithmic increment of the particle grid. By defining $R_p$ and $\epsilon_{{\rm I},p}$ as for the photons, it yields:
\begin{equation}
\Delta p(p,\omega)/p_{\rm{min}}(p,\omega) > \epsilon_{\rm{I},p}  ~.\label{eq_I_particles}
\end{equation}
Contrary to the photon case, there is no simple equation for the boundary defining this region and the implicit equation \ref{eq_I_particles} must be solved numerically. The boundary $\omega_{\rm I}(p)$ is computed easily since there is only one positive solution to the equation on $\omega$ and it happens to be always smaller than $\omega_0^*(p)$. A simple approximation for this boundary, accurate in all regimes to better than 30$\%$, is: 
\begin{equation}
\omega_{\rm I}(p) \approx \frac{\epsilon_{{\rm I},p} p/2}{2p^2+(1+\epsilon_{{\rm I},p}-\sqrt{1+\epsilon_{{\rm I},p}})p+\epsilon_{{\rm I},p}+2} ~.
\end{equation}
When the resolution is good enough ($R_\omega>>1$, $\epsilon_{{\rm I},p} << 1$), condition \ref{eq_I_particles} reduces to that used in Eq. 78 of \citet{Poutanen08} for electron down-scattering: $\omega >  \epsilon_{{\rm I},p}/(4\gamma)$  for relativistic particles. 

\begin{figure}[h!]
\begin{center}
\includegraphics[width=\columnwidth]{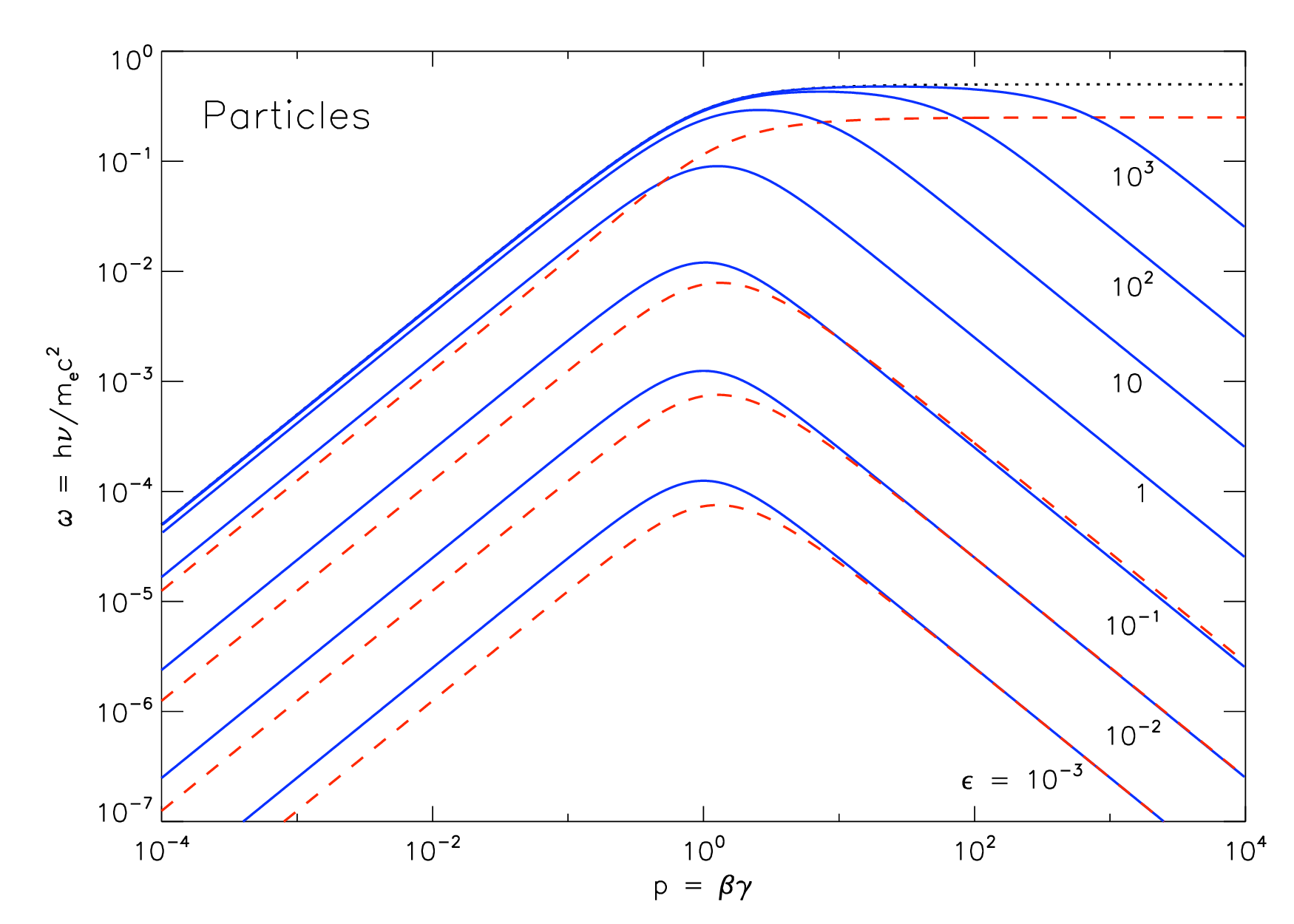}
\caption{Particle equation: boundaries $\omega_{{\rm I}}(p)$ for the integral approach (solid lines) and $\omega_{{\rm FP}}(p)$ for the Fokker-Planck approximation (dashed lines) for $\epsilon_{{\rm I},p}=10^{-3},10^{-2},10^{-1},1,10,10^2,10^3$ and $\epsilon_{\rm FP}=10^{-3},10^{-2},10^{-1},1$ respectively. The integral approach and the FP approximation are valid respectively above and below the related boundaries. The dotted line shows $\omega_0^*(p)$.}
\label{fig_bdr_part}
\end{center}
\end{figure}

Fig. \ref{fig_bdr_part} shows this boundary for various resolutions ($R_p \approx 9200,930,97,13,3.8,2.0,1.3$ and $n=5$).
Contrary to the equation for photons, solving the evolution equation for particles with a simple numerical integration requires very high resolution to guarantee a correct accuracy. For example, soft photons of energy $\omega=10^{-5}$ scattering off high energy particles ($\gamma=10^2$) are accurately described by the integral approach only if $R_p > 900$, which is far beyond current desktop computer capabilities and the constraint is even more severe when mildly relativistic particles are involved. 

\subsubsection{Accuracy}
\begin{figure}[h!]
\begin{center}
\includegraphics[width=\columnwidth]{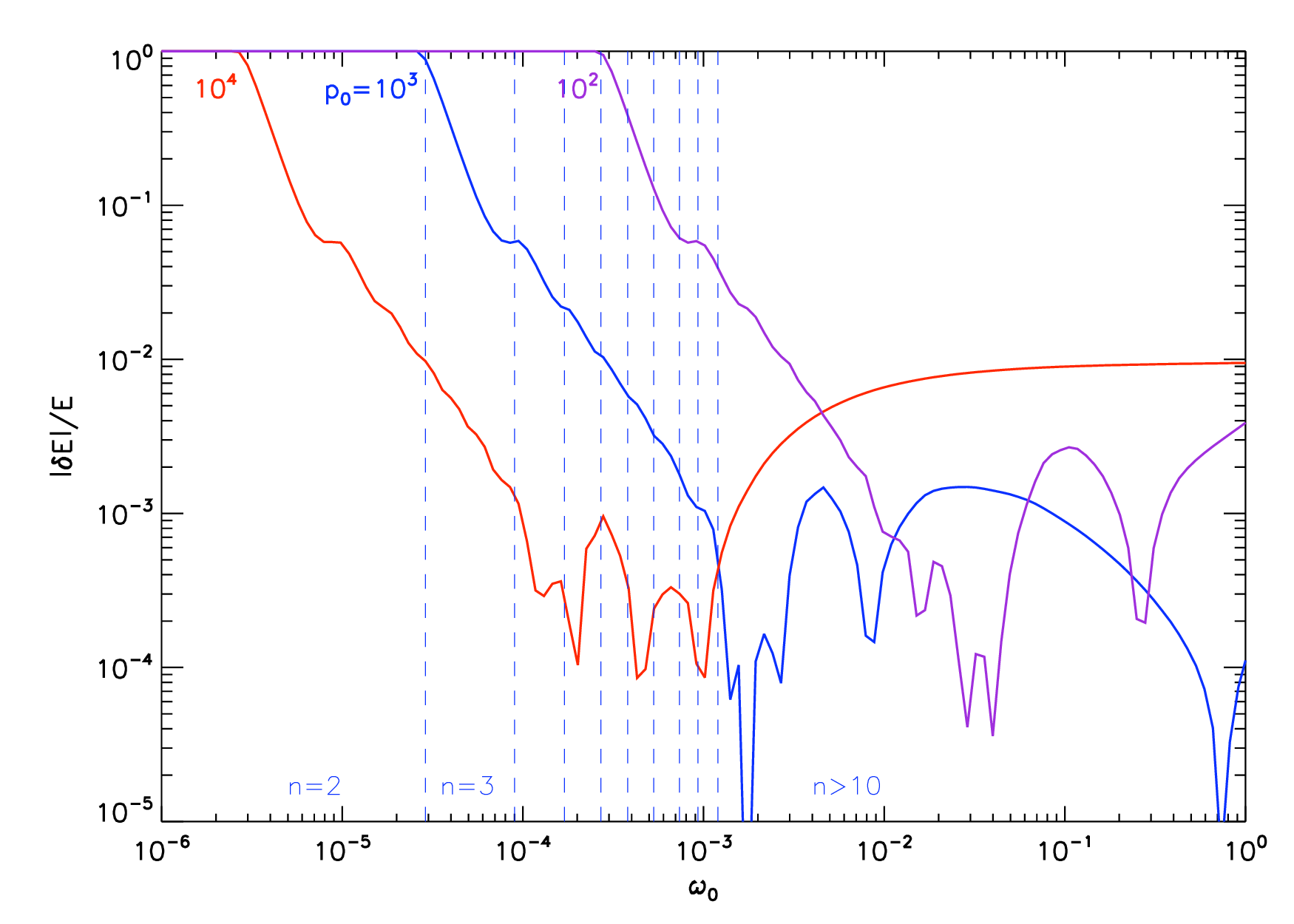}
\caption{Numerical error on energy conservation, for a single time step, as a function of the incident photon energy $\omega_0$, for particles of momentum $p_0=10^2,10^3,10^4$. In the case $p_0=10^3$, the boundaries between zones where the scattered distribution spans 2,3,4,5,6,7,8,9, and at least 10 energy bins are over-plotted (vertical dashed lines). The error is computed as the relative error between the total energy losses of the photon population and the total energy losses of the electron population: $\delta E /E = (|\partial_t E_p| - |\partial_t E_\omega|)/(|\partial_t E_p| + |\partial_t E_\omega|)$. For this run the resolutions are: $R_p=64/6$ and $R_\omega=256/12$.} \label{int_prec}
\end{center}
\end{figure}

If conditions \ref{eq_bdr_I_photons} and \ref{eq_I_particles} are not satisfied over the entire simulation domain, numerical computation of the Compton scattering may lead to inaccuracy. In particular, energy is not conserved when logarithmic grids are used  (as explained before), and its measure provides a good way to estimate numerical errors\footnote{For linear grids, other accuracy indicators should be used.}. 
Fig. \ref{int_prec} shows the error on energy when soft mono-energetic photons are up-scattered by high energy particles. The numerical computation was performed with the code presented in \citet{Belmont08}, on a single time step, using an explicit scheme on logarithmic energy grids, and turning off all processes/injection/escape but Compton scattering. The contribution of Compton scattering was forced to be computed by the integral approach for the equation on particles. The contribution to the equation on photons was also computed with the integral approach but given the energy range and the grid resolution considered here, this method is accurate in this case. Both initial distributions were set to zero except in one energy bin, and they were normalised to unity to keep the number of scattering events constant as the photon energy is varied. The error was measured by computing the total energy lost by particles $\partial_t E_p$ and comparing it to the energy gained by photons $\partial_t E_\omega$.  

For low energy photons, the scattering takes place far below the boundary in the $(p,\omega)$ plane of Fig. \ref{fig_bdr_part}. The scattered distribution typically spans less than 2 bins. The energy change of particles is not captured at all by the numerical scheme and the error is 100\%. As the photon energy increases, the scattered distribution widens and the error decreases. From the cases presented here, we find that an error less than 1\% corresponds to scattered distributions that span more than 5 bins. When $n>10$, the error is dominated by other weak numerical errors.

By the choice of particle and photon energies, we have only focused here on the error made in the equation for particles. A similar study can be done for photons. However, it would lead to very similar results and contrary to the case of particles, integration errors for photons are less relevant to astrophysical applications.

\subsection{Fokker-Planck approach}
Alternatively, Compton scattering can be described by the Fokker-Planck method. This approximation assumes that the cross section and the {\it initial} particle (or photon) distribution are only weakly dependent on the energy of the incoming particle (or photon) on the scale of the scattered distribution width. When the width of the scattered distribution is small, a second order Taylor expansion of the integral equation can be performed, which gives the well-known Fokker-Planck equations:
\begin{eqnarray} 
\partial_t N_\omega &=& \partial_\omega \left(A_\omega N_\omega\right) + \frac{1}{2} \partial^2_{\omega^2} \left(D_\omega N_\omega\right) \\
\partial_t N_e &=& \partial_p \left(\frac{\gamma}{p} A_\gamma N_e\right) + \frac{1}{2} \partial_{p} \left( \frac{\gamma}{p} \partial_p \left(\frac{\gamma}{p} D_e N_e\right)\right)
\end{eqnarray}
with 
\begin{eqnarray}
A_\omega ; D_\omega &=& \int_0^{\infty}  c \bar{\sigma}_{1;2}(p;\omega) dN_e(p) \\ 
A_p ; D_p &=& \int_0^{\infty} c \bar{\sigma}_{1;2}(p;\omega) d N_\omega(\omega) ~.
\end{eqnarray}

As for the integral approach, the validity of the FP approximation depends on the energy of the incoming particle and photon. Although the real criteria should in principle depend on the initial distributions and should be computed at each time step of a time dependant simulation, it is more convenient to define an approximate but general criteria. 
%Hereafter we assume that the energy scale for the variation of the cross section and the particle energy is of the order of the kinetic energy	$\partial_{\gamma_0} \ln{\left[N_e(p_0)\sigma(\gamma_0,\omega0;\omega)\right]} \approx 1/ (\gamma_0-1)$ and $\partial_{\omega_0} \ln{\left[N_\omega(\omega_0)\sigma(\gamma_0,\omega0;\omega)\right]} \approx 1/ \omega_0$.

\subsubsection{Particles}
The FP method for particles assumes that the quantity $N_e(\gamma_0)d\bar{\sigma}(\gamma_0,\omega_0;\omega)$ does not vary much with $\gamma_0$ in the width $\Delta \gamma$ of the scattered distribution. There is no unique way to define such a condition. By assuming that the typical energy scale for the variation of this quantity is the initial kinetic energy: $\gamma_0-1$, then the FP approach is found to give a correct description of the particle evolution if: 
\begin{equation}
\Delta\gamma(\omega,\gamma)/(\gamma-1) < \epsilon_{\rm FP} \label{eq_ppFP0}
\end{equation}
where $\epsilon_{\rm FP}<<1$ is a small parameter that describes the accuracy of the FP approximation: the smaller $\epsilon_{\rm FP}$ the more valid the FP approach. The equation for the boundary of this region of the $(p,\omega)$ space reads:
\begin{equation}
\omega < \omega_{{\rm FP}}(p)
\end{equation}
where
\begin{equation}
\omega_{{\rm FP}} = \frac{\gamma}{2}\left( \sqrt{\frac{1-\epsilon_{\rm FP}(1-1/\gamma)\beta}{1-\epsilon_{\rm FP}(1-1/\gamma)/\beta}}-1\right) ~.
\end{equation}
In the limit $\epsilon_{\rm FP} << 1$, a simple approximation for all particle energy is:
\begin{eqnarray}
\omega_{{\rm FP}} \approx \frac{\epsilon_{\rm FP}\beta}{4(\gamma+1)} ~,
\end{eqnarray}
which reduces to $\omega_{{\rm FP},p} \sim \epsilon_{\rm FP}/(4\gamma)$ in the ultra-relativistic regime ($p >> 1$). Boundaries $\omega_{\rm FP}(p)$ are shown in Fig. \ref{fig_bdr_part} for $\epsilon_{\rm FP}=10^{-3},10^{-2},10^{-1}$ and 1. The Fokker-Planck approximation for particles is a good approximation when very low energy photons are up-scattered by mid-relativistic particles. It fails easily when the particle energy becomes large. For example, by setting $\epsilon_{\rm FP}=0.1$, the scattering of photons off particles with energy $p=10^3$ can only be described by the FP method when the photon energy is less than $\omega=10^{-5}$. However, when one does not need such a precision and sets $\epsilon_{\rm FP} \la 1$, the FP method is valid over a large range of energy.

\subsubsection{Photons}
Similarly, the FP method for photons assumes that the quantity $N_\omega(\omega_0)d\bar{\sigma}(\gamma_0,\omega_0;\omega)$ does not vary much with $\omega_0$ in the width $\Delta\omega$ of the scattered distribution. By assuming that the typical energy scale for the variation of this quantity is of the order of the incoming photon energy $\omega_0$, the FP approach for photons is valid if:
\begin{equation}
\Delta\omega(\omega,\gamma) /\omega < \epsilon_{\rm FP} ~. \label{eq_wpFP0}
\end{equation}
When solved as a function of $\omega$, the equation for the boundary is again the solution of a second order polynomial and the condition \ref{eq_wpFP0} is satisfied in the region defined by: 
\begin{equation}
p< p^c_{\rm FP} \quad \rm{and} \quad  \rm{min}\left[\omega_{\rm FP}^c(p),\omega_{\rm FP}^-(p)\right] < \omega < \omega_{\rm FP}^+(p)\\
\end{equation}
where
\begin{eqnarray} 
p^c_{\rm FP} &=&  \frac{(4-\epsilon_{\rm FP}) - (4+\epsilon_{\rm FP})\sqrt{1-\epsilon_{\rm FP}}}{4\epsilon_{\rm FP}} ~. \\
\omega^\pm_{\rm FP} &=& \frac{1}{4} \left[ 2(1-p-\gamma) + \epsilon_{\rm FP}(\gamma+p) \right. \nonumber \\ 
  & \pm & \left. \sqrt{\epsilon_{\rm FP}}\sqrt{4(1-\gamma-p)+\epsilon_{\rm FP}(1+2p\gamma+2p^2)} \right] ~, \\
\omega^c_{\rm FP}&=& \frac{p}{\epsilon_{\rm FP}} \times \nonumber \\ 
&& \left(1-\frac{\epsilon_{\rm FP}^2/2/p^2}{1+\epsilon_{\rm FP}/\beta+\sqrt{(1+\epsilon_{\rm FP}/\beta)^2-\epsilon_{\rm FP}^2/p^2} } \right) ~.
\end{eqnarray}
As for the integral approach, the inverse equation for this boundary  $p_{{\rm FP}}(\omega)$ is more practical to use and is best evaluated numerically. 

Boundaries $p_{\rm{FP}}$ are plotted in Fig. \ref{fig_photons} for $\epsilon_{\rm FP }=10^{-3},10^{-2},10^{-1}$, and 1. Contrary to the equation for particles, the photon evolution is only poorly described by the FP approximation and only low energy photons ($\omega<0.1$) scattering off low energy particles ($p<0.1$) undergo small angle scattering that can be accurately described by the FP approximation.

%Contrary to the integral method the FP approximation always conserved energy even when used in regimes where it fails so that check on the total conservation of energy is not a appropriate methods to estimate errors. 

\subsection{Combined approach}
As can be seen in Fig. \ref{fig_photons} and \ref{fig_bdr_part}, the two methods happen luckily to give a correct description of Compton scattering in regions of the $(p,\omega)$ space that can be complementary. This suggests that combining the two methods is a good way to overcome numerical accuracy issues in computing the effects of Compton scattering. However, this can only be done if the validity regions for the two methods cover the entire simulation domain, that is, for given $\epsilon_{{\rm I},p}$, $\epsilon_{{\rm I},\omega}$ and $\epsilon_{\rm FP }$, if:
\begin{eqnarray}
p_{{\rm FP}}(\omega) \ge p_{{\rm I}}(\omega) &&\quad \forall \omega ~, \label{eq_overlap_photons} \\
 \omega_{{\rm FP}}(p) \ge \omega_{{\rm I}}(p) &&\quad \forall p \label{eq_overlap_particles}
\end{eqnarray}
for the equation on photons and particles respectively. 
Typically, this implies that $\epsilon_{{\rm I},\omega} \le \epsilon_{\rm FP}$ and $\epsilon_{{\rm I},p} \le \epsilon_{\rm FP}$, which sets a constraint on the grid resolution:
\begin{equation}
R_{\omega;p} \ge \frac{n-1}{\epsilon_{\rm{FP}}}  ~.
\end{equation}
As good accuracy for each methods requires $n>5$ and $\epsilon_{\rm{FP}}<<1$, this puts strong constraints on the resolution. For example, by setting $\epsilon_{\rm FP}=0.1$ and $n=5$ this leads to $R>40$. For particle grids that typically span 5 orders of magnitude, it implies the need to use 200-bin grids accessible to most desktop computers. However, for photon grids that typically span 15 orders of magnitude, it corresponds to 600-bin grids, which requires high computing power.

\begin{figure}[h!]
\begin{center}
\includegraphics[width=\columnwidth]{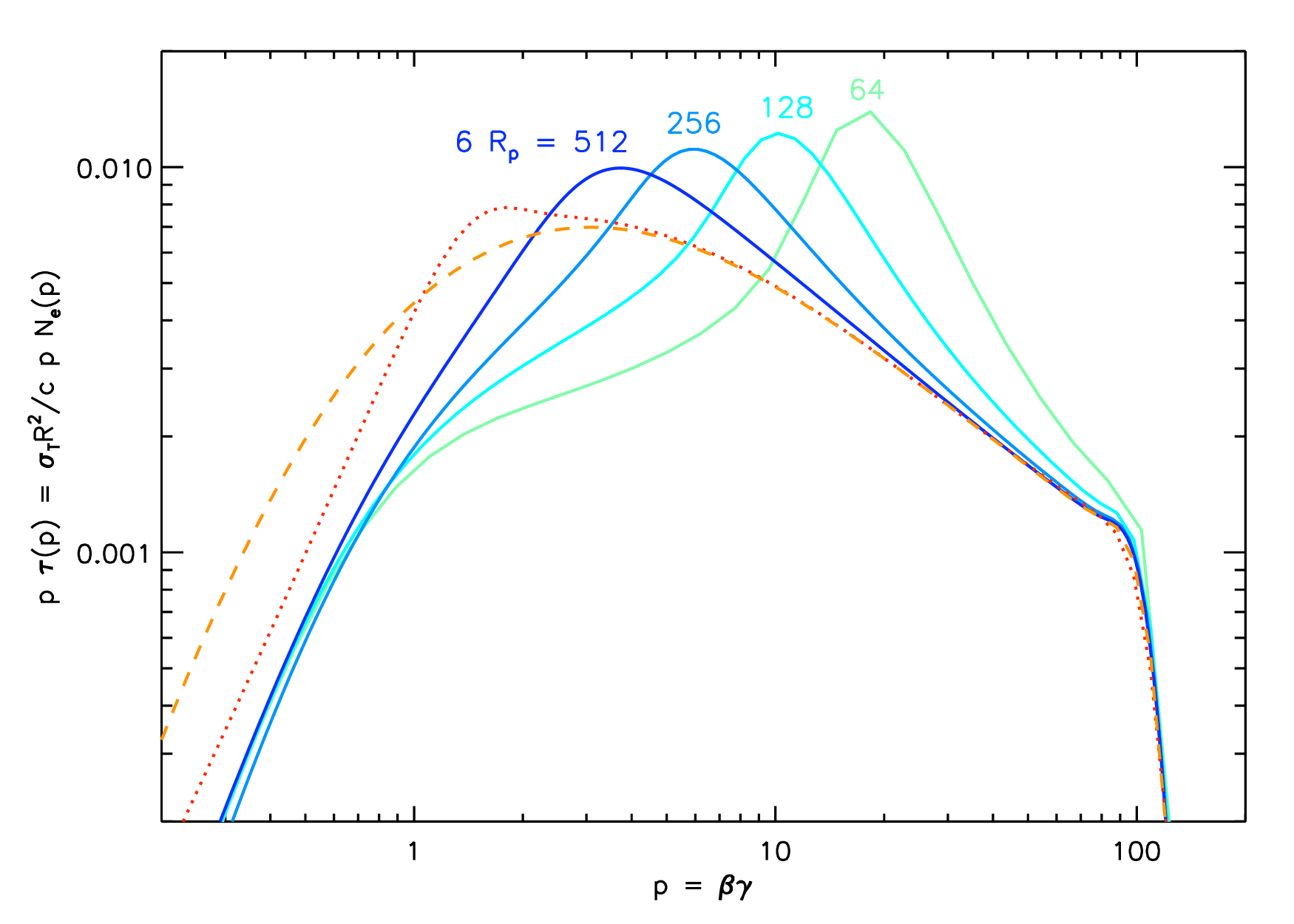}
\includegraphics[width=\columnwidth]{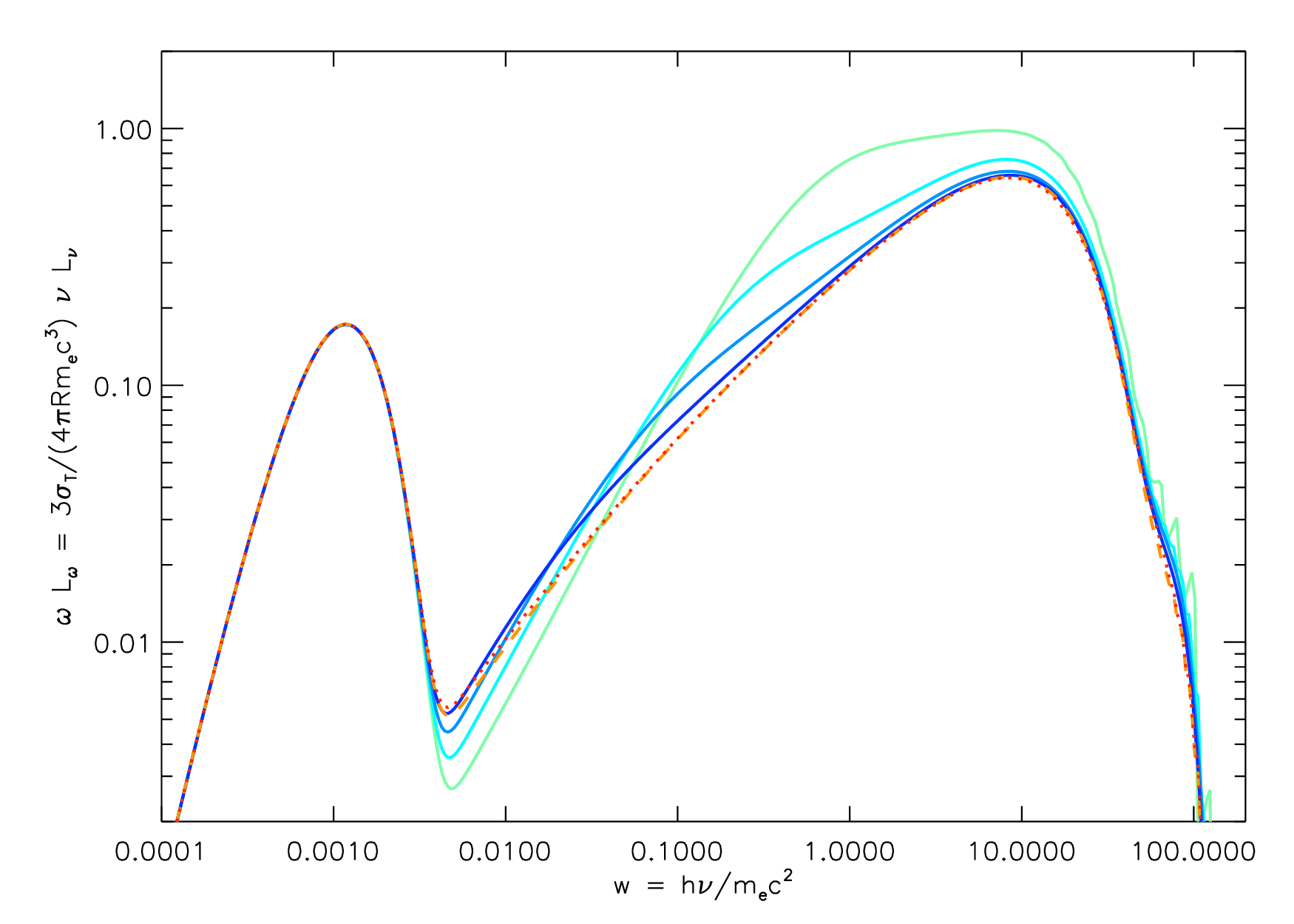}
\caption{Particle distribution and spectra in steady state when Compton scattering for particles is treated with the integral method for different resolutions: $R_p=512/6, 256/6, 128/6$, and $64/6$ (solid lines), when it is treated with the combined approach with $\epsilon_{\rm FP}=0.1$ and $n=10$ (dashed line), and with the Fokker-Planck approximation (dotted line). For these runs, the power injected in particles and photons is: $l_{\rm in, e}=10$, $l_{{\rm in},\omega}=1$ and the source size is set to $R=10^7$ cm. The temperature of the seed photons is $k_BT/m_ec^2=3\times10^{-4}$ and the energy of the injected particles is $\gamma=10^2$. For the combined and FP approaches, the particle resolution is:  $R_p=64/6$. For all runs, the resolution of the photon grid is $R_\omega=512/9$. } \label{fig_steady}
\end{center}
\end{figure}

For each equation (for photons and for particles), one can define an average boundary when the validity regions of both methods overlap:
\begin{equation}
\omega_c(p) = \left( \omega_{\rm{I}} \omega_{{\rm FP}} \right)^{1/2} \quad{\rm and}\quad p_c(\omega) = \left(p_{\rm{I}} p_{{\rm FP}} \right)^{1/2}~.
\end{equation}
Finally, a combination of the two methods is achieved by solving the following equations:
\begin{eqnarray}
\partial_t N_{e}(p) &=&  \partial_p \left( \frac{\gamma}{p} A^{\omega<\omega_c}_e N_e \right)  \nonumber \\
             &+& \frac{1}{2}\partial_p \left( \frac{\gamma}{p} \partial_p \left(\frac{\gamma}{p}D_e^{\omega<\omega_c} N_e \right) \right) \nonumber \\  
                                            &+& \int dN_e(p_0) \int_{\omega_c(p_0)}^{\infty}  c \frac{d\bar{\sigma}}{dp}(p_0,\omega_0; \omega(p)) dN_\omega(\omega_0)  \nonumber  \\ 
                                            &- & N_{e}(p) \int_{\omega_c(p)}^{\infty}  c \bar{\sigma}_0(\omega_0,p)dN_\omega(\omega_0)  ~, \\
\partial_t N_\omega(\omega) &=&   \partial_\omega\left( A^{p<p_c}_\omega N_\omega \right) +\frac{1}{2} \partial^2_{\omega^2} \left(D_\omega^{p<p_c} N_\omega \right) \\ 
&+& \int dN_\omega(\omega_0) \int_{p_c(\omega_0)}^{\infty}  c \frac{d\bar{\sigma}}{d\omega}(p_0,\omega_0 ; \omega) dN_e(p_0)   \nonumber \\ 
                                                    & - &   N_\omega(\omega) \int_{p_c(\omega)}^{\infty}  c \bar{\sigma}_0(\omega,p_0) dN_e(p_0)  
\end{eqnarray}
where integrations in the integral part are performed only above the boundary energy $\omega_c$ and momentum $p_c$ and where the FP coefficients are computed as integrals of the cross section only below these boundaries:
\begin{eqnarray}
A_\omega ; D_\omega &=& \int_0^{p_c(\omega)}c \sigma_{1;2}(p;\omega) d N_e(p)  \\ 
A_p ; D_p &=& \int_0^{\omega_c(p)} c \sigma_{1;2}(p;\omega) d N_\omega(\omega) ~.
\end{eqnarray}
Fig. \ref{fig_steady} shows the steady photon and particle distributions for a more realistic simulation. Particles are injected with a mono-energetic distribution at high energy ($\gamma=10^2$) at a constant rate (the compactness of which is set to $l_{\rm in,e}=\sigma_T P /(m_ec^3R)=10$, where $P$ is the injected power) and escape at a constant rate with the speed of light to allow for a steady state. Soft photons are also injected with a black body spectrum of temperature $k_BT=3\times10^{-4} m_ec^2$ and a flux of compactness $l_{{\rm in},\omega}=\sigma_T L_{\rm soft} /(m_ec^3R)=1$  \citep[see e.g.][for detailed definitions of the compactness parameters]{Belmont08}. We start with an empty system and the code simultaneously evolves both distributions with time until they reach a steady state. The only process turned on is Compton scattering. Fig. \ref{fig_steady} shows the results of three kinds of runs where the contribution of Compton scattering to the evolution of the particle distribution was forced to be treated with the integral (several resolutions), the Fokker-Planck and the combined approaches. The distributions look very different. 

As expected, the integral approach fails to capture efficiently the particle cooling for low resolution runs: the lower the resolution, the steeper the slope of the high energy tail of the particle distribution. These excess high energy particles more efficiently up-scatter the soft photons and the resulting spectra are harder. 
In steady state, the total energy loss must balance the total injected power. Power is supplied through particles $l_{{\rm in},e}=10$ and through photons: $l_{\rm in}=l_{\rm in,e}+l_{{\rm in},\omega}=11$. Energy is also lost both through particles and photons. The steady particle distribution has low density $\tau=\sigma_TR\int dN_e =2.3\times10^{-2}$ so that the energy lost through particle escapes is rather small ($l_{\rm out,e}\approx 1$) and most of the energy escape as radiation. As numerical errors prevent particles from cooling as they should, the overall system actually gains energy artificially and the total energy loses are measured to be larger than the injected power: $l_{\rm out}  = l_{\rm out,\omega}+l_{\rm out,e} =19.6, 14.2, 12.2$, and $11.4$ from lower to higher grid resolution. The corresponding effective electron temperatures are $k_BT_e=6.23, 4.68, 3.95$, and $3.60~m_ec^2$ \citep[see Eq. 2.8, ][]{Coppi92}.

The Fokker-Planck equation provides good energy conservation and the plasma luminosity balances the injected power: $l_{\rm out}=11.0$. However, energy conservation does not guarantee accurate results and significant deviation is observed in the low energy part of the particle distribution, where the Fokker-Planck approximation is expected to fail. In the particular case presented here, this has no effect on the emitted spectrum since the spectrum is dominated by the Comptonisation by high energy particles. Nevertheless it could strongly affect the predicted spectrum when other processes such as Coulomb collisions or pair annihilation are involved. The electron temperature is measured to be: $k_BT_e=3.40~m_ec^2$.

The combined approach produces the best results. With low grid resolution ($R_p=64/6$), it gives the correct photon spectrum and the correct particle distribution, it conserves energy to better than $1\%$ accuracy ($l_{\rm out}=11.0$) and it was checked that the shape of the steady distributions does not depend on the grid resolution (not shown here). The electron temperature is $k_BT_e=3.39~m_ec^2$. In comparison, the integral approach requires more than 10 times higher resolution to produce similar results, which also means at least 10 times more computational time. Extensive tests show that the combined approach is very efficient in most cases. It takes advantage of the two approaches. The way the two methods are associated is numerically simple to implement and it can easily be shown that the total number of particles and photons as well as the the total energy are conserved. Moreover even when conditions \ref{eq_overlap_photons} and \ref{eq_overlap_particles} are not exactly satisfied this method still provides good accuracy in many cases since, as long as these conditions are not strongly violated, only a small faction of particles and/or photons is treated inaccurately. Also, even if the evolution of a large fraction of particles and/or photons is not described correctly by the numerical scheme they might have only a minor contribution to the overall distribution evolution in some specific cases. For these reasons some simulations might reach a correct final accuracy, despite violating the accuracy condition. However, this is very problem-dependent and the accuracy should be checked very carefully when these conditions are not satisfied.

\section*{Conclusion}
In this paper we have investigated numerical issues related to the computation of isotropic Compton scattering in numerical codes. We have derived a form of the distribution of Compton scattered photons that is free of numerical cancellations. This form is exact with no limitation on the photon and electron energies. The numerical accuracy resulting from the direct use of the cross section has been studied for various grid resolutions and equations have been proposed for the boundaries where this method must be replaced by an approximate Fokker-Planck treatment to guarantee sufficient accuracy. All results presented here have been included in and extensively tested with the new code developed by \citet{Belmont08}.

\begin{acknowledgements}
The author thanks J. Poutanen for providing the numerical routines corresponding to the formulae of NP94. 
\end{acknowledgements}

\bibliographystyle{aa}

\end{document}